\documentclass[11pt,epsf,epsfig]{article}
\usepackage{graphics}
\usepackage{graphicx,epsfig}					
\usepackage{amssymb}		
\usepackage{amsmath}

\newcommand{\gsim}{\lower.7ex\hbox{$\;\stackrel{\textstyle>}{\sim}\;$}}
\newcommand{\lsim}{\lower.7ex\hbox{$\;\stackrel{\textstyle<}{\sim}\;$}}

\def\stilde{\widetilde}

\newcommand{\newc}{\newcommand}
\newc{\Nc}{N_{c}}
\newc{\CG}{C_G}
\newc{\gp}{g'}
\newc{\stopi}{\stilde t_i}
\newc{\sboti}{\stilde b_i}
\newc{\staui}{\stilde \tau_i}
\newc{\stopj}{\stilde t_j}
\newc{\sbotj}{\stilde b_j}
\newc{\stauj}{\stilde \tau_j}
\newc{\stopI}{\stilde t_1}
\newc{\stopII}{\stilde t_2}
\newc{\sbotI}{\stilde b_1}
\newc{\sbotII}{\stilde b_2}
\newc{\stauI}{\stilde \tau_1}
\newc{\stauII}{\stilde \tau_2}
\newc{\sstop}{s_{t}}
\newc{\cstop}{c_{t}}
\newc{\ssbot}{s_{b}}
\newc{\csbot}{c_{b}}
\newc{\sstau}{s_{\tau}}
\newc{\cstau}{c_{\tau}}
\newc{\Sstop}{s_{2t}}
\newc{\Cstop}{c_{2t}}
\newc{\Ssbot}{s_{2b}}
\newc{\Csbot}{c_{2b}}
\newc{\Sstau}{s_{2\tau}}
\newc{\Cstau}{c_{2\tau}}
\newc{\salpha}{s_\alpha}
\newc{\calpha}{c_\alpha}
\newc{\Calpha}{c_{2\alpha}}
\newc{\Salpha}{s_{2\alpha}}
\newc{\sbetapm}{s_{\beta_\pm}}
\newc{\cbetapm}{c_{\beta_\pm}}
\newc{\Sbetapm}{s_{2 \beta_\pm}}
\newc{\Cbetapm}{c_{2 \beta_\pm}}
\newc{\sbetaO}{s_{\beta_0}}
\newc{\cbetaO}{c_{\beta_0}}
\newc{\SbetaO}{s_{2 \beta_0}}
\newc{\CbetaO}{c_{2 \beta_0}}
\newc{\vu}{v_u}
\newc{\vd}{v_d}
\newc{\seL}{\stilde e_L}
\newc{\smuL}{\stilde \mu_L}
\newc{\seR}{\stilde e_R}
\newc{\smuR}{\stilde \mu_R}
\newc{\suL}{\stilde u_L}
\newc{\sdL}{\stilde d_L}
\newc{\suR}{\stilde u_R}
\newc{\sdR}{\stilde d_R}
\newc{\scL}{\stilde c_L}
\newc{\ssL}{\stilde s_L}
\newc{\scR}{\stilde c_R}
\newc{\ssR}{\stilde s_R}
\newc{\snue}{\stilde \nu_e}
\newc{\snumu}{\stilde \nu_\mu}
\newc{\snutau}{\stilde \nu_\tau}
\newc{\Gpm}{G^\pm}
\newc{\Hpm}{H^\pm}
\newc{\FFbS}{\overline{FF}S}
\newc{\FFbV}{\overline{FF}V}
\newc{\FSS}{F_{SS}}
\newc{\FSSS}{F_{SSS}}
\newc{\FFFS}{F_{FFS}}
\newc{\FFFbS}{F_{\overline{FF}S}}
\newc{\FSSV}{F_{SSV}}
\newc{\FVS}{F_{VS}}
\newc{\FVVS}{F_{VVS}}
\newc{\FFFV}{F_{FFV}}
\newc{\FFFbV}{F_{\overline{FF}V}}
\newc{\Fgauge}{F_{\rm gauge}}
\newc{\DRbarprime}{$\overline{\rm DR}'$ }
\newc{\DRbar}{$\overline{\rm DR}$ }
\newc{\MSbar}{$\overline{\rm MS}$ }
\newc{\Yu}{{\bf Y}_u}
\newc{\Yd}{{\bf Y}_d}
\newc{\Ye}{{\bf Y}_e}
\newc{\Au}{{\bf a}_u}
\newc{\Ad}{{\bf a}_d}
\newc{\Ae}{{\bf a}_e}
\newc{\bm}{{\bf m}}

\newc{\rwino}{r_{\tilde W}}
\newc{\rmu}{r_{\tilde H}}
\newc{\ra}{r_A}

\newcommand{\nc}{\newcommand}

\nc{\beaa}{\begin{eqnarray*}} \nc{\eeaa}{\end{eqnarray*}}
\nc{\beq}{\begin{equation}}   \nc{\eeq}{\end{equation}}
\nc{\bea}{\begin{eqnarray}}   \nc{\eea}{\end{eqnarray}}
\nc{\baa}{\begin{array}}      \nc{\eaa}{\end{array}}
\nc{\bit}{\begin{itemize}}    \nc{\eit}{\end{itemize}}
\nc{\ben}{\begin{enumerate}}  \nc{\een}{\end{enumerate}}
\nc{\bce}{\begin{center}}     \nc{\ece}{\end{center}}
\nc{\non}{\nonumber}

\def\bed{\begin{description}}
\def\eed{\end{description}}

\def\non{\nonumber}

\def\k1slash{k_1\hspace{-10.5pt}/\ \ }

\def\simge{\mathrel{%
   \rlap{\raise .57ex \hbox{$>$}}{\lower .57ex \hbox{$\sim$}}}}
\def\simle{\mathrel{
   \rlap{\raise 0.512ex \hbox{$<$}}{\lower 0.512ex \hbox{$\sim$}}}}

\begin{document}
\setlength{\baselineskip}{0.22in}

\title{\textbf{The LHC Discovery Potential of a Leptophilic Higgs}}

\author{Shufang Su\footnote{shufang@physics.arizona.edu} \,and\, Brooks Thomas\footnote{brooks@physics.arizona.edu}\\
\\
\it{Department of Physics, University of Arizona, Tucson, AZ
85721}}

\date{}
\maketitle

\begin{abstract}
In this work, we examine a two-Higgs-doublet
extension of the Standard Model in which one Higgs doublet is 
responsible for giving mass to both up- and down-type quarks, while a 
separate doublet is responsible for giving mass to leptons.  
We examine both the theoretical and experimental constraints on 
the model and show that large regions of parameter space are allowed
by these constraints in which
the effective couplings between the lightest neutral Higgs scalar
and the Standard-Model leptons are substantially enhanced.
We investigate the collider phenomenology
of such a ``leptophilic'' two-Higgs-doublet model and show that in cases 
where the low-energy spectrum contains only one light, \(CP\)-even scalar, 
a variety of collider processes essentially irrelevant for the discovery of a
Standard Model Higgs boson (specifically those in which the Higgs 
boson decays directly into a charged-lepton pair) can contribute 
significantly to the discovery potential of a light-to-intermediate-mass
($m_h\lesssim140$~GeV) Higgs boson at the LHC.
\end{abstract}

\section{Introduction\label{se:intro}}

One of the primary goals of the Large Hadron Collider (LHC), a proton-proton 
collider with a center of mass energy \(\sqrt{s}=14\)~TeV, will 
be to investigate the sector responsible for the breaking of the electroweak symmetry.  
In the Standard Model (SM), a single Higgs doublet is responsible for the
spontaneous breakdown of the \(SU(2)_L\times U(1)_Y\) gauge group to \(U(1)_{EM}\).
The coupling constants of the sole physical Higgs scalar to the rest of the SM particles 
are completely determined by their masses, and consequently there is little guesswork
involved in determining the most promising channels~\cite{Asai:2004ws,Abdullin:2005yn} 
in which one might hope to discover such a scalar.  For a relatively light
($114\mbox{~GeV}\lesssim m_h\lesssim 125\mbox{~GeV}$) SM Higgs boson, 
those channels are 
$gg\rightarrow h\rightarrow\gamma\gamma$ and $t\bar{t}h(h\rightarrow b\bar{b})$, while  
for an intermediate-mass ($125\mbox{~GeV}\lesssim m_h\lesssim 140\mbox{~GeV}$) Higgs, 
the single most promising channel is the weak-boson fusion (WBF)~\cite{wbf} process 
$qq'\rightarrow qq'h(h\rightarrow\tau\tau)$~\cite{Rainwater:1998kj}.  
For a heavier Higgs, with  
$ m_h\gtrsim 140\mbox{~GeV}$, the most relevant channels are 
$h\rightarrow WW^{\ast}$ 
and $h\rightarrow ZZ^{\ast}$, with the Higgs produced via either 
gluon fusion or WBF~\cite{Asai:2004ws,Abdullin:2005yn}.

In models where the Higgs sector differs significantly from that of the 
Standard Model, however, the situation can change dramatically.  This is 
true even in cases where the low-energy effective theory describing a given model at 
the weak scale contains only a
single, light, $CP$-even Higgs scalar. 
Indeed, at low energies, many models with extended Higgs sectors 
have effective descriptions that are ``Standard-Model-like'' in 
the sense that they contain a single light Higgs boson, but one whose 
couplings to the Standard Model fermions and gauge bosons differ --- 
potentially significantly --- from those of a SM Higgs.  
Such discrepancies, in turn, can 
translate into vast differences in LHC phenomenology: some (or, in severe 
cases, even all) of the standard detection channels
for a SM Higgs may disappear as a result of such modifications, while 
others, related to processes buried beneath background in the SM, may 
become crucial for discovery.  

One set of channels which are not terribly significant for the discovery of a SM Higgs, but
could become so in models with modified Higgs sectors, consists of those involving direct decays of
the Higgs boson to a pair of high-$p_T$ leptons.  In the SM, a light Higgs boson 
(with mass $m_h<130$~GeV) decays predominantly into $b\bar{b}$, and the ratio 
$\mathrm{BR}(h\rightarrow \ell\ell)/\mathrm{BR}(h\rightarrow b\bar{b})$ 
(where $\ell = e,\mu,\tau$) is roughly proportional to $m_\ell^2/m_b^2$,
due to the fact that in the SM, the same Higgs doublet is responsible for giving mass to 
both quarks and leptons.   Consequently, attention has been
focussed predominately on processes in which the Higgs boson decays to a tau pair
(with a branching ratio of about 10\%), and in particular on 
the weak-boson fusion process
$qq'\rightarrow qq'h(h\rightarrow\tau\tau)$.  This is the only 
process particularly relevant for SM Higgs discovery in which the Higgs 
decays directly to leptons, though it is 
now regarded as one of the most promising 
discovery channels for a SM Higgs in the intermediate mass 
region~\cite{wbf,ATLASTDR,CMSTDR}.   Searching for the Higgs in the 
$gg\rightarrow h \rightarrow \tau\tau$ and 
$tth(h\rightarrow \tau\tau)$ channels is more difficult, 
due to a combination of factors, including 
enhanced SM backgrounds and suppressed signal cross-sections.  
 
By contrast, processes in which a SM Higgs boson decays into first- or 
second-generation leptons are generally assumed to be irrelevant for discovery. 
This is because under the assumption of Yukawa-coupling universality among the 
lepton generations (an assumption we will be making throughout the present work), 
the small size of $m_{\mu}$ compared to $m_\tau$ results in 
${\rm BR}(h\rightarrow \mu\mu)$ being roughly two orders of magnitude smaller than 
${\rm BR}(h\rightarrow \tau\tau)$, with ${\rm BR}(h\rightarrow ee)$ nearly three 
orders of magnitude smaller still.  Consequently, the rates for processes involving
$h\rightarrow \mu\mu$ and $h\rightarrow ee$ are extremely suppressed relative to those
involving tau pairs, both in the SM 
and in most simple extensions of the Higgs sector.    
On the other hand, there are strong motivations for considering processes of this
sort at the LHC.  Experimentally, a signal involving a pair of high-$p_{T}$ muons or electrons will be easy to identify, as the muon- and electron-identification efficiencies at each 
of the LHC detectors are each greater than 90\%\cite{ATLASTDR, CMSTDR}.  
Furthermore, once a Higgs boson 
is discovered in these channels, its mass could be readily reconstructed with high 
precision.  Such channels could also be of use in determining the Higgs Yukawa 
couplings to leptons.  

Two-Higgs-doublet models (2HDM), which stand as perhaps the simplest, most 
tractable example of a non-minimal electroweak-symmetry-breaking sector, provide
a useful context in which to study the role of leptonic Higgs-decay processes.
These models arise in a number of beyond-the-Standard-Model contexts from supersymmetry 
to little Higgs scenarios~\cite{ArkaniHamed:2001nc} and have a rich 
phenomenology, many of whose consequences for LHC physics are still being 
uncovered.  In general, 2HDM can be categorized according to how the Higgs doublets 
couple to the SM quarks and leptons.  In what has become known as a Type~I 2HDM, one 
doublet is responsible for the masses of both quarks and leptons, while the other 
decouples from the fermions entirely.  In a Type~II 2HDM, one Higgs doublet couples to the
up-type quark sector, while the other Higgs doublet couples to both the down-type quark sector and the charged leptons --- as is the case, for example, in the Minimal Supersymmetric Standard Model (MSSM).  In both of these standard scenarios, the leptonic branching 
ratios for a light Higgs do not differ much from their SM values throughout most of
parameter space\footnote{There are, however, regions of parameter space in the MSSM 
within which the effective $hb\bar{b}$ coupling is suppressed due 
to radiative corrections~\cite{HaberKQ}, and consequently 
${\rm BR}(h\rightarrow \ell\ell)$ becomes large.}, since the same doublet 
gives masses to both the bottom quark and the charged leptons. 

One interesting alternative possibility, which will be the primary 
focus of the present work, is a 2HDM
scenario in which one Higgs doublet couples exclusively to (both up- and down-type) 
quarks, while the other couples exclusively to leptons --- a scenario which we 
will henceforth dub the leptophilic two-Higgs-doublet model (L2HDM)\footnote{In the literature, this
scenario has also been referred to as the lepton-specific 2HDM~\cite{Barger:2009me},
leptonic 2HDM~\cite{Goh:2009wg}.}.  
This model has been discussed 
previously in the literature in relation to its effect on Higgs branching fractions 
and decay widths~\cite{Barger:2009me,Barnett:1983mm,Barnett:1984zy,Barger:1989fj}, 
flavor physics~\cite{Grossman:1994jb}, and potential implications for neutrino 
phenomenology~\cite{Aoki:2008av} and dark matter studies~\cite{Goh:2009wg}.  
Some analyses of the LHC phenomenology of the model were presented in Ref.~\cite{Goh:2009wg}, which focused on the non-decoupling region of the parameter space where
additional physical Higgs scalars are light.

In this work, 
we discuss the leptonic decays of the lightest $CP$-even Higgs scalar 
in the L2HDM at the LHC.  In particular, we examine the 
discovery potential in a decoupling regime in which only
one light scalar, which resembles the SM Higgs, appears in the low-energy effective 
description of the model.  
We begin in Section~\ref{sec:Param} by presenting the model and reviewing how the 
coupling structure of the lightest neutral Higgs particle is modified from that of a
SM Higgs.     
In Section~\ref{sec:Bounds}, we discuss the applicable experimental constraints from flavor physics, direct searches, etc.\ and show that they still permit substantial deviations in
the couplings between the Higgs boson and the other SM fields away from their Standard-Model
values.
In Section~\ref{sec:BRXSec}, we discuss the implications of
such modifications on the Higgs branching ratios and production rates.  
In Section~\ref{sec:LHCSignatures}, we discuss potential Higgs discovery channels in which
the Higgs boson decays directly into a pair of charged leptons, and in Section~\ref{sec:Combined}, we calculate the discovery potential for a light, 
leptophilic Higgs using the combined results from all of these leptonic channels.
In Section~\ref{sec:Conclusion} we conclude.

\section{The Leptophilic 2HDM\label{sec:Param}}

The L2HDM, as defined here, is a modification of the 
SM in which the Higgs sector consists of two \(SU(2)_L\times U(1)_Y\) 
scalar doublets, both of which receive nonzero vacuum expectation values.  The first 
of these doublets, which we call $\phi_q$, couples only to 
(both up- and down-type) quarks, while the other, which we call $\phi_{\ell}$, 
couples only to leptons.  In other words, the Yukawa interaction Lagrangian is
specified to be
\begin{equation}
  \mathcal{L}_{\mathit{Yukawa}}=-(y_u)_{ij}\bar{q}_i\phi_q^c u_j -
    (y_d)_{ij}\bar{q}_i\phi_q d_j -(y_e)_{ij}\bar{\ell}_i\phi_\ell e_j+h.c.,
    \label{eq:Yuks}
\end{equation}
where $(y_u)_{ij}$, $(y_d)_{ij}$, and $(y_e)_{ij}$ are $3\times 3$ Yukawa 
matrixes, $q_i$ and $\ell_i$ respectively denote the left-handed quark and lepton 
fields, $u_i$ and $d_i$ respectively denote the right-handed up- and down-type
quark fields, and $e_i$ denotes the right-handed lepton fields.    
This coupling structure can be achieved by imposing a 
\(\mathbb{Z}_2\) symmetry under which \(\phi_\ell\) and \(e_i\) are odd, while all 
the other fields in the model are even. 
We will assume that this symmetry is broken only softly,  
by a term of the form \((m_{q\ell}^2\phi_q^{\dagger}\phi_\ell+h.c.)\)
in the scalar potential.  

In the L2HDM, that scalar potential takes the usual form
common to all two-Higgs doublet models.
Assuming that there is no \(CP\)-violation in the Higgs 
sector, this potential can be parameterized as follows~\cite{Gunion:2002zf}:    
\begin{eqnarray}
V&=& m_{1}^2|\phi_q|+m_{2}^2|\phi_\ell|^2
  +\left(m_{q\ell}^2\phi_q^{\dagger}\phi_\ell+h.c.\right)
  \nonumber\\&&+
  \lambda_{1}(|\phi_q|^2)^2+
  \lambda_{2}(|\phi_\ell|^2)^2
  +\lambda_{3}|\phi_q|^2|\phi_\ell|^2
  +\lambda_{4}|\phi_q^{\dagger}\phi_\ell|^2+
  \frac{\lambda_5}{2}\left[(\phi_q^\dagger \phi_\ell)^2+h.c.\right]\label{eq:Vstandard}
\end{eqnarray} 
It is assumed that the parameters of the theory are assigned such that both $\phi_q$
and $\phi_\ell$ acquire nonzero VEVs (which we respectively denote $v_q$ and $v_\ell$),
and that $v_q^2+v_\ell^2=v^2\equiv(174~\mathrm{GeV})^2$.     
We define $\tan\beta$ as 
\begin{equation}
  \tan\beta\equiv v_q/v_\ell, 
  \label{eq:DefBeta}
\end{equation}
so that large $\tan\beta$ corresponds to small $v_\ell$, and therefore to large 
intrinsic lepton Yukawa couplings. 
In the broken phase of the theory, the spectrum of the model includes the three 
massless Goldstone modes which become the longitudinal modes of the $W^\pm$ and $Z$ bosons,
as well as five massive scalar degrees of freedom: two $CP$-even fields $h$ and $H$, a pseudoscalar $A$, and a pair of charged fields $H^\pm$.  The relationship between 
the physical $CP$-even Higgs scalars $h$, $H$   and the real, neutral
degrees of freedom in $\phi_q$ and $\phi_\ell$ is parameterized by the mixing 
angle $\alpha$:
\begin{equation}
  \left(\begin{array}{c} H\\ h\end{array}\right)=\sqrt{2}\left(
  \begin{array}{cc} \cos\alpha & \sin\alpha \\ -\sin\alpha & \cos\alpha\end{array}\right)
  \left(\begin{array}{c} \mathrm{Re}[\phi_{\ell}^0-v_{\ell}]\\
  \mathrm{Re}[\phi_q^0-v_q]\end{array}\right).
  \label{eq:Defalpha}
\end{equation} 
In what follows, we will focus primarily on the physics of $h$, the lightest of these two
scalars.

Since the potential given in Eqn.~(\ref{eq:Vstandard}) includes eight model parameters 
 --- $\lambda_i$ $(i=1,\ldots,5)$, $m^2_1$, $m^2_2$, and $m_{ql}^2$ ---  which are subject 
to the constraint $v_{q}^2+v^2_\ell = (174\ {\rm GeV})^2$,  seven of these eight
parameters may be considered free.  In what follows, it will be useful to work in
a different, more physically meaningful basis for these parameters:  
\begin{equation}
(m_h, m_H, m_A, m_{H^\pm}, \tan\beta, \sin\alpha, \lambda_{5}), 
\label{eq:MeaningfulBasis}
\end{equation}
where $m_h$, $m_A$, $m_H$, and $m_{H^\pm}$ are the masses of the corresponding physical Higgs 
scalars. 

In order to study the collider phenomenology of the L2HDM,
it will be necessary to characterize how the effective couplings between $h$ 
and the SM fields differ from their 
SM values.
Eqn.~(\ref{eq:Yuks}) indicates that the effective couplings between the 
fermions and \(h\) are given
in terms of these mixing angles\footnote{Note that these expressions depend on the 
conventions~(\ref{eq:DefBeta}) and~(\ref{eq:Defalpha})
used in defining $\alpha$ and $\beta$, and hence frequently differ from source to 
source within the literature.} by  
\begin{eqnarray}
   \mathcal{L}_{\mathit{h\bar{f}f}}&=&
  - \frac{m_u}{\sqrt{2}v} \frac{\cos\alpha}{\sin\beta}h\bar{u}_Lu_R-
   \frac{m_d}{\sqrt{2}v} \frac{\cos\alpha}{\sin\beta}h\bar{d}_Ld_R+
   \frac{m_{e}}{\sqrt{2}v}\frac{\sin\alpha}{\cos\beta}h\bar{e}_Le_R+h.c.
   \label{eq:HffCouplings}.
\end{eqnarray}
Following~\cite{Phalen:2006ga}, we can define a set of parameters \(\eta_i\) which represent
the ratios of these effective couplings to their SM values.  At tree level,
\begin{equation}
  \eta_{u}=\eta_{d}=\frac{\cos\alpha}
  {\sin\beta}~~~~,~~~~\eta_{\ell}=-\frac{\sin\alpha}{\cos\beta}.
\end{equation}
Similarly, one can also define \(\eta\)-parameters for the trilinear couplings of
$h$ with the electroweak gauge bosons, with the result that
\begin{equation}
  \eta_W=\eta_Z\equiv \eta_V =\sin(\beta-\alpha).
\end{equation}

Since a certain set of effective couplings whose leading contributions occur at one
loop --- namely \(hgg\) and \(h\gamma\gamma\) --- are also relevant to the collider phenomenology of 
Higgs bosons, it is worth deriving \(\eta\)-factors for them as well.  
The effective operators that give rise to \(hgg\) and \(h\gamma\gamma\)
are~\cite{Gunion:1989we}
\begin{equation}
  \left(\sum_q\eta_qF_{1/2}(\tau_q)\right)\frac{h}{\sqrt{2}v}
  \frac{\alpha_3}{8\pi}G^a_{\mu\nu}G^{a\mu\nu},\label{eq:GluonEffLoop}
\end{equation}
\begin{equation}
  \left(\eta_WF_1(\tau_W)+3\sum_q Q_q^2\eta_qF_{1/2}(\tau_q)
  +\sum_{\ell}\eta_{\ell}F_{1/2}(\tau_{\ell})\right)\frac{h}{\sqrt{2}v}
  \frac{\alpha}{8\pi}F_{\mu\nu}F^{\mu\nu},
\end{equation}
where $\tau_i=4m_i^2/m_h^2$, $Q_q$ is the electric charge of quark $q$, and
\begin{eqnarray}
  F_{1/2}(\tau)&=& -2\tau[1+(1-\tau)f(\tau)]\label{eq:F12def} \\
  F_1(\tau)& =& 2+3\tau+3\tau(2-\tau)f(\tau)
\end{eqnarray}
and
\begin{equation}
  f(\tau)=\left\{ \begin{array}{cc} {\rm arcsin}^2(1/\sqrt{\tau}) & \tau\geq 1 \\
  -\frac{1}{4}\left[ \log (\eta_+/\eta_-) -i\pi\right]^2 & \tau<1
  \end{array}\right.
  \end{equation}
  with $\eta_\pm =(1\pm \sqrt{1-\tau})$.  When $F_1(\tau_i)$ and
$F_{1/2}(\tau_i)$ are complex (which occurs when $m_h>2m_i$), it corresponds to internal
lines going on shell.  
This allows us to define a scaling factor for each of these effective vertices:
\begin{eqnarray}
\eta_{g}&=&\frac{\sum_q\eta_qF_{1/2}(\tau_q)}{\sum_qF_{1/2}(\tau_q)}~=~\eta_q\label{eq:EtaGandEtaQ}\\
  \eta_{\gamma}&=& \frac{\eta_W F_1(\tau_W)+3\sum_q Q_q^2 \eta_qF_{1/2}(\tau_q)
  +\sum_{\ell}\eta_{\ell}F_{1/2}(\tau_{\ell})}{F_1(\tau_W)+3\sum_q Q_q^2 F_{1/2}(\tau_q)
  +\sum_{\ell}F_{1/2}(\tau_{\ell})},
 \end{eqnarray} 
Since \(F_{1/2}(\tau_f)\) has an overall \(m_{f}^2\) prefactor (from the \(\tau_f\)), 
the contribution from top quarks running in the loops will still dominate over the
contribution from leptons unless \(\eta_\ell/\eta_q\sim 10^4\); thus the lepton loops
generally can be neglected.  It is worth noting that
since the effective Higgs-gluon-gluon coupling receives contributions solely from quark
loops, \(\eta_g=\eta_q\) to leading order in \(\alpha_s\), whereas 
\(\eta_{\gamma}\) depends on \(\eta_q\),
\(\eta_{\ell}\), and \(\eta_{W}\) in a nontrivial way.

The mixing angles $\alpha$ and $\beta$ are constrained by several theoretical
consistency conditions, as well as a number of experimental constraints.  We
will put off discussion of the latter until Section~\ref{sec:Bounds} and focus
on the former.  First of all, we require that the Higgs sector not be strongly
coupled, in the sense that all $\lambda_i$ may be considered perturbatively small 
(i.e.\ $\lambda_i < 4\pi$ for all $i=1,\ldots,5$) and that the $S$-matrix satisfies all 
relevant tree-unitarity constraints. 
This implies that the quartic couplings $\lambda_i$ appearing in
Eqn.~(\ref{eq:Vstandard}) must satisfy~\cite{Unitarity}
\begin{eqnarray}
\frac{1}{2}\left(3(\lambda_1+\lambda_2)\pm\sqrt{9(\lambda_1-\lambda_2)^2+4(2\lambda_3+\lambda_4|)^2}\right)<8\pi
~~,~~~~ \lambda_3+2\lambda_4\pm|\lambda_5|<8\pi\nonumber\\
\frac{1}{2}\left(\lambda_1+\lambda_2\pm\sqrt{(\lambda_1-\lambda_2)^2+4|\lambda_5|^2}\right)<8\pi
~~,~~~~~~~~\lambda_3\pm\lambda_4 < 8\pi~~~~~~~~~~~\nonumber\\
\frac{1}{2}\left(\lambda_1+\lambda_2\pm\sqrt{(\lambda_1-\lambda_2)^2+4|\lambda_5|^2}\right)<8\pi
~~,~~~~~~~~\lambda_3\pm|\lambda_5|<8\pi.~~~~~~~~~~
\label{eq:PertBounds}
\end{eqnarray}
Perturbativity constraints also apply to 
the Yukawa couplings $y_u$, $y_d$, and $y_e$ appearing in Eqn.~(\ref{eq:Yuks}), 
which are modified from their SM values according to 
Eqn.~(\ref{eq:HffCouplings}).  However, since $y_u$ and $y_d$ are suppressed relative
to their SM values  rather than enhanced when $\tan\beta>1$ (the case of interest here), 
no stringent constraints arise on
account of such modifications.  In addition to these perturbativity constraints,
we must also require that the scalar potential given in Eqn.~(\ref{eq:Vstandard}) is 
finite at large field values and contains no flat directions.  These considerations
translate into the bounds~\cite{Gunion:2002zf}
\begin{equation}
  \lambda_{1,2}>0~~~,~~~\lambda_3>-2\sqrt{\lambda_1\lambda_2}~~~,~~~
  \lambda_3+\lambda_4-|\lambda_5|> -2\sqrt{\lambda_1\lambda_2}.
  \label{eq:VacStabBounds}
\end{equation}

In this work, we will be primarily interested in examining situations in which
the additional physical scalars $H^\pm$, $H$, and $A$ are heavy enough to ``decouple''
from the collider phenomenology of the theory in the sense that the only observable 
signals of beyond-the-Standard-Model physics at the LHC  at low luminosity  involve the light $CP$-even scalar $h$.  For 
our present purposes, it will be sufficient to define our ``decoupling regime'' by the
condition that $m_{H^\pm},m_H,m_A>M$, where $M$ is some high scale.  
Of course this regime includes 
the strict decoupling limit in which $M\rightarrow\infty$ and the
mixing angles satisfy the condition $\alpha\approx\beta-\pi/2$.  However, 
it also includes substantial regions of parameter space within which the values of 
$\alpha$ and $\beta$ deviate significantly from this relationship.  

\begin{figure}[ht!]
  \begin{center}
  \epsfxsize 3.5 truein \epsfbox{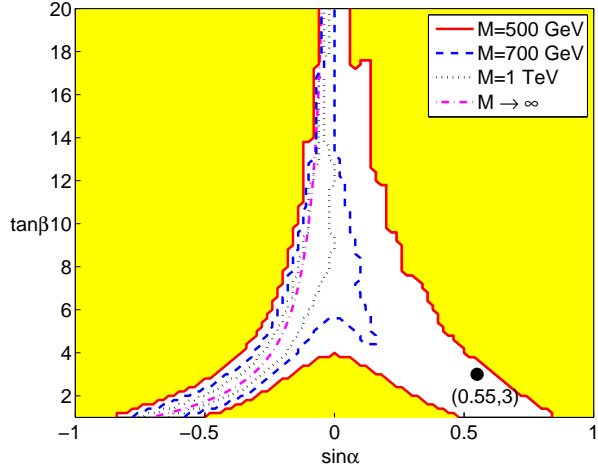}
  \end{center}
  \caption{The decoupling region of $\sin\alpha$ - $\tan\beta$ parameter space within which all
  perturbativity and vacuum-stability constraints are simultaneously satisfied.  The
  three contours shown correspond to $m_{H^\pm},m_A,m_H> M$ for $M=500$~GeV (solid line), $M=700$~GeV (dashed line),
  and $M=1000$~GeV (dotted line).  The pure decoupling limit in which $m_{H^\pm},m_A,m_H\rightarrow\infty$
  is indicated by the dash-dotted line.  The dot marks the point 
  $( \sin\alpha = 0.55, \tan\beta = 3 )$, which will be used as a benchmark point 
  in the analysis presented in Sections~\ref{sec:BRXSec} and~\ref{sec:LHCSignatures}.
  Within the shaded region, at least one of scalars $H$, $A$ or $H^\pm$ is light ($<500$ GeV).
  \label{fig:SinaTanbExcl}}
\end{figure}

The extent of parameter space allowed according to our definition of the
decoupling regime is illustrated in Fig.~\ref{fig:SinaTanbExcl}. 
This figure shows the decoupling regions of $\sin\alpha$ - $\tan\beta$ parameter space 
in which all of the aforementioned constraints are satisfied for a variety
of different values of $M$.  Contours corresponding to $M=500$~GeV, $M=700$~GeV, 
and $M=1$~TeV are displayed, along with a dash-dotted line representing the pure decoupling
limit, where $m_{H^\pm},m_H,m_A\rightarrow\infty$ and $\alpha\approx\beta-\pi/2$.   
The contours in Fig.~\ref{fig:SinaTanbExcl} were obtained by 
fixing $m_h$ to a particular value ($120$~GeV) and surveying over 
the remaining parameters.  A given combination of
$\sin\alpha$ and $\tan\beta$ is considered to be ``allowed'' in this sense as long as 
there exists some combination of model parameters for which 
$m_{H^\pm},m_H,m_A>M$, and for which all of the 
constraints in Eqs.~(\ref{eq:PertBounds}) 
and~(\ref{eq:VacStabBounds}) are simultaneously satisfied.
It is readily apparent from the figure that
sizable regions of parameter space exist within which all constraints are satisfied, yet
the masses of all scalars other than $h$ are large enough to effectively decouple from
the low-energy effective description of the model.  It is also apparent that for
$M\gg 1$~TeV, the decoupling region, as we have defined it, approaches the pure
decoupling limit. 

\begin{figure}[ht!]
  \epsfxsize 2.1 truein \epsfbox{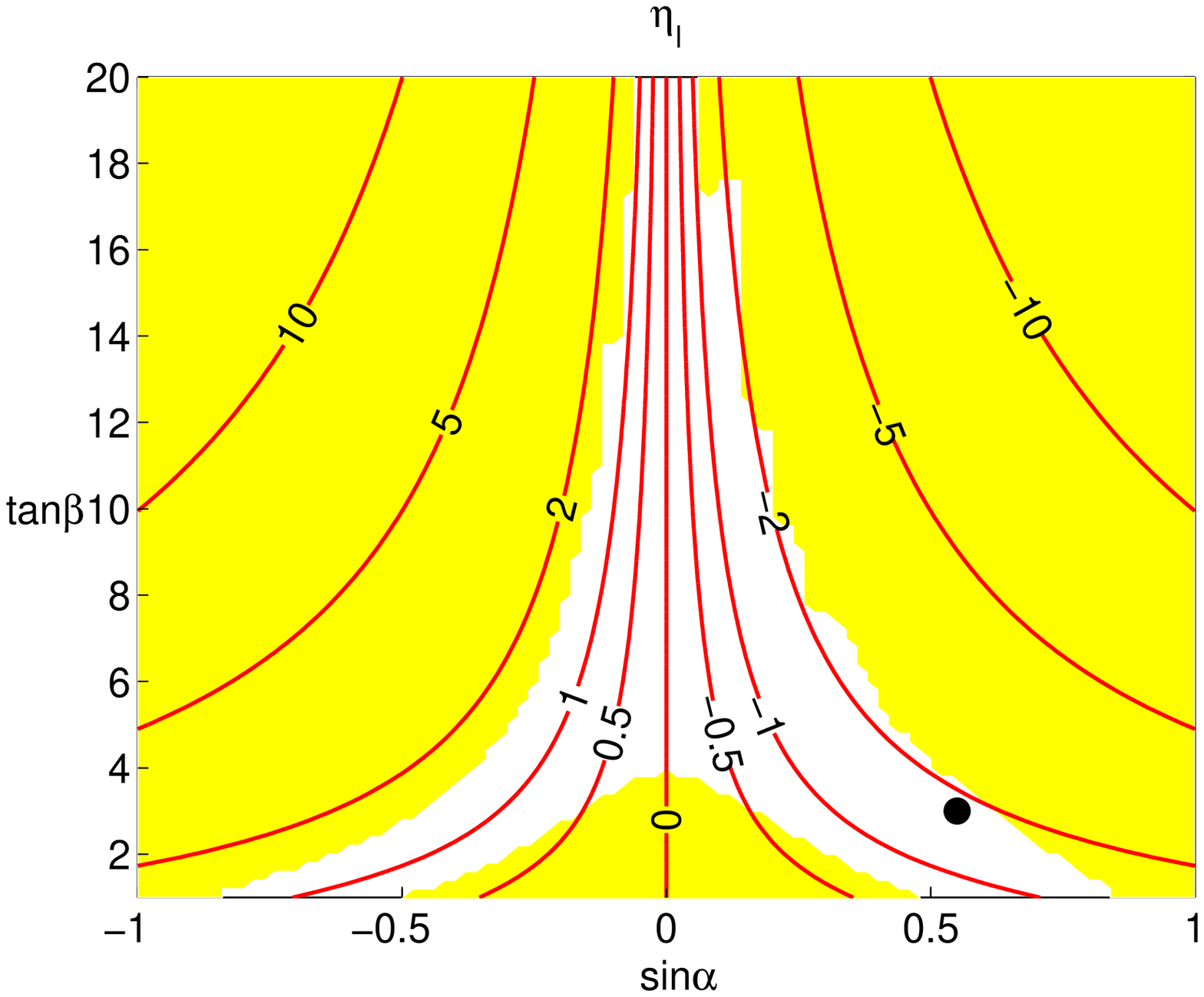}
  \epsfxsize 2.1 truein \epsfbox{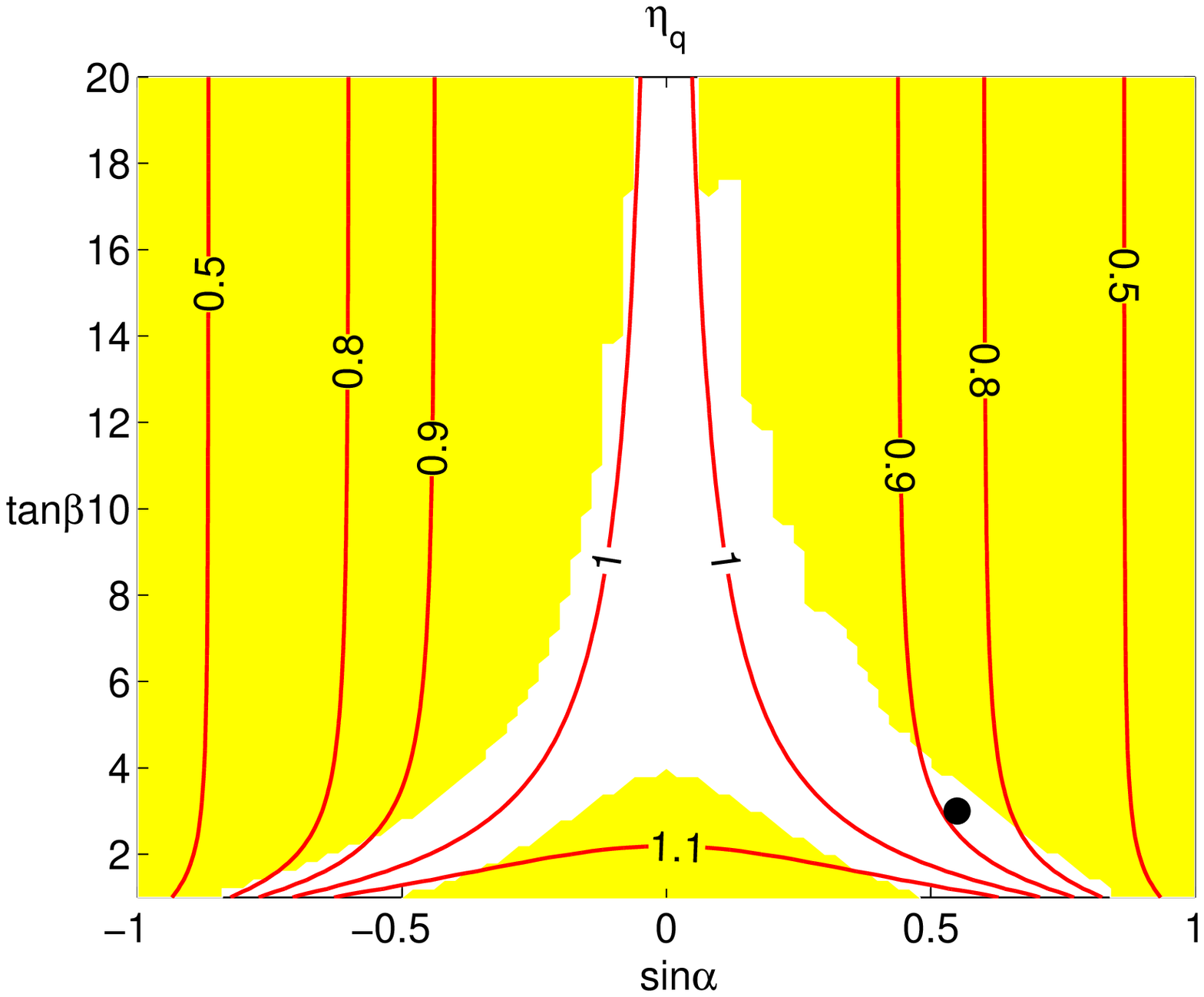}
  \epsfxsize 2.1 truein \epsfbox{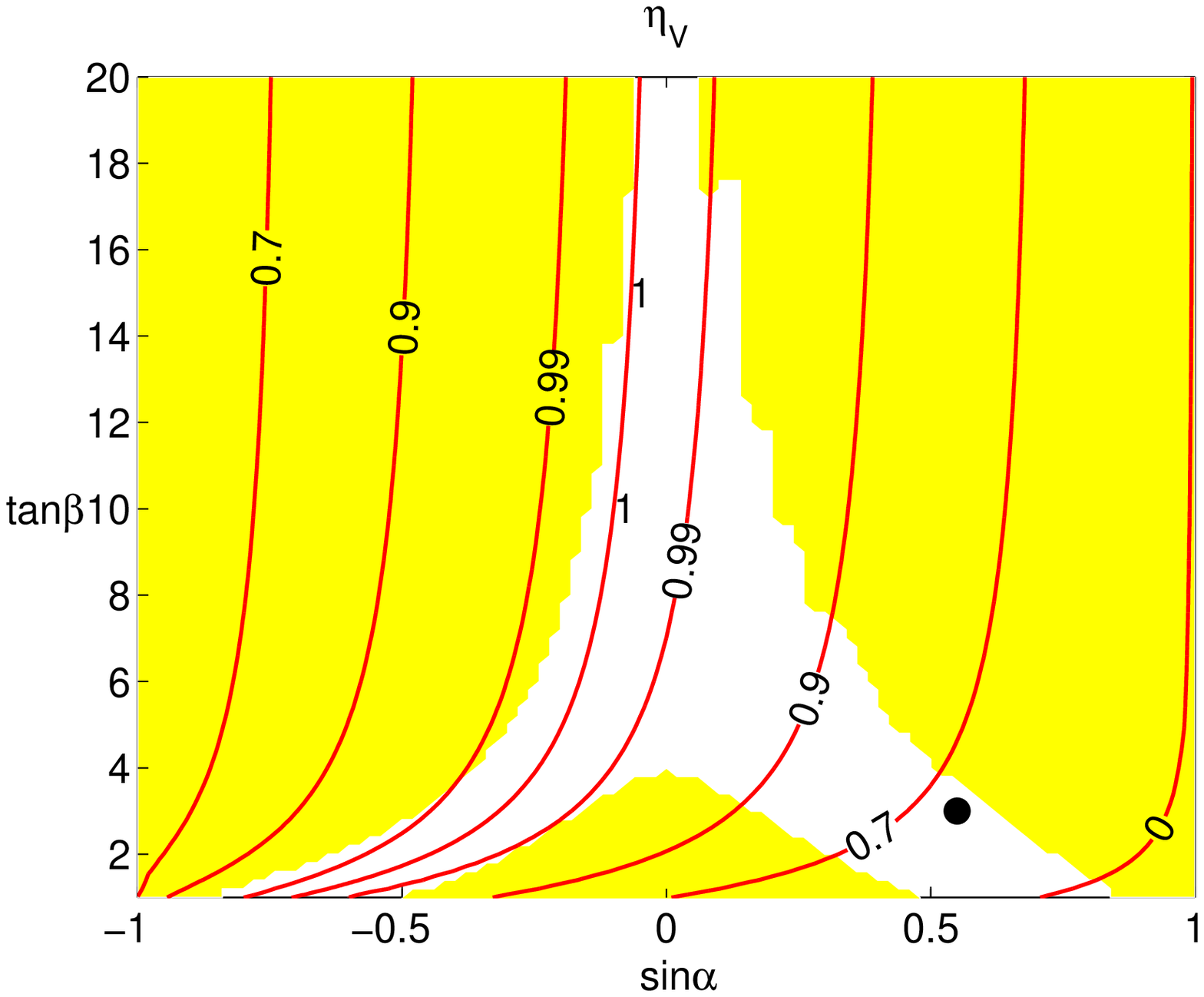}
  \caption{ 
  Contours for $\eta_\ell$ (left), $\eta_q$(center), and $\eta_V$(right) in $\sin\alpha$ $-$ $\tan\beta$ space in the L2HDM. 
  Superimposed on each of these panels is an outline of the region within 
  which all perturbativity and vacuum-stability constraints are simultaneously satisfied 
  for $m_{H^\pm},m_H,m_A>500$~GeV, as in Fig.~\ref{fig:SinaTanbExcl}.
  The dot marks the benchmark point 
  $( \sin\alpha = 0.55, \tan\beta = 3 )$.
  \label{fig:SinaTanbEtaCont}}
\end{figure}

It is interesting to inquire
to what extent the effective Higgs couplings can 
be modified in the decoupling regime without running afoul of the 
aforementioned constraints.  In the three panels shown in Fig.~\ref{fig:SinaTanbEtaCont} we
plot a number of contours in $\sin\alpha$ - $\tan\beta$ parameter space 
corresponding to different values of $\eta_\ell$ (left), $\eta_q$ (center)
and $\eta_V$ (right).  On each panel, we have also superimposed the $M = 500$~GeV
 contour from Fig.~\ref{fig:SinaTanbExcl}.  It is evident from these
plots that while $\eta_\ell,\eta_q,\eta_V \rightarrow 1$ in the $M\rightarrow\infty$
limit, large regions of parameter space are still allowed in the
decoupling regime within these $\eta$-factors can deviate substantially from unity.      
The message, then, is that the effective couplings of a light Higgs boson in the 
decoupling regime do not have to approximate those which correspond to the pure 
decoupling limit, in which they approach those of a SM Higgs.  On the contrary, a wide 
variety of possibilities for the mixing angles $\alpha$ and $\beta$ are still open 
in this regime, and as we shall soon see, many of these possibilities have a
dramatic effect on in the collider phenomenology of the scalar $h$.

\section{Experimental Constraints\label{sec:Bounds}}

In addition to the perturbativity and vacuum-stability bounds discussed in the previous
section, the L2HDM is constrained by  additional considerations related to flavor-physics experiment, direct searches, etc.   
We now proceed to investigate these constraints in an effort to show that they can easily
be satisfied in the decoupling regime --- even in the
region of parameter space most interesting for collider phenomenology, where $\tan\beta$ is large and $\sin\alpha$ deviates substantially from zero.

Let us begin with those bounds related to direct searches for 
beyond-the-Standard-Model scalars at LEP.
The current direct detection bounds (at 95\% CL) for the masses of charged and 
neutral $CP$-odd Higgs bosons, as reported by the particle data group~\cite{Yao:2006px}, 
are $m_{H^{\pm}}\geq 78.6 \mathrm{~GeV}$ and $m_{A}\geq 93.4 \mathrm{~GeV}$.  
These clearly present no problem for the model in the decoupling limit 
considered here.  

Far more 
stringent constraints on models with more than one Higgs doublet can be derived, however,
from experimental limits on flavor-violating processes that receive
contributions at the one-loop level from diagrams involving charged Higgs bosons.
Let us first consider flavor violation in the lepton sector, which is
constrained by analyses of \(\tau\rightarrow \mu\gamma\), 
\(\mu\rightarrow e\gamma\), \(\tau\rightarrow \mu ee\), and
\(\mu\rightarrow e\) conversion in nuclei.  In the 
absence of neutrino masses, the matrix of effective $H^+\bar{\nu}_i e_j$ couplings is 
proportional to the charged-lepton mass matrix; hence there is no additional source
of lepton-flavor violation (LFV).  In the presence of neutrino masses this is
no longer true, and nonzero contributions to LFV processes arise at one loop 
due to diagrams with charged Higgs bosons running in the loop.  However, it has
been shown~\cite{Grimus:2002ux} that the resultant flavor-violating amplitudes are
several orders of magnitude below current experimental bounds.  Therefore,  
even in cases in which the effective $H^+\bar{\nu}e^-$ coupling receives a substantial 
$\tan\beta$-enhancement factor, such sources of LFV will not present any
phenomenological difficulties.    

Now let us turn to consider the situation in the quark sector, where, by contrast,
flavor-violation rates can be sizeable.  
This is because
the effective $H^+\bar{u}_i d_j$ couplings in two-Higgs-doublet
models include flavor-violating terms proportional to the off-diagonal elements of the 
Cabibbo-Kobayashi-Maskawa (CKM) matrix $V_{ij}$: 
\begin{equation}
\mathcal{L}_{H^\pm \bar{f}f^\prime} =- \frac{\cot\beta}{v}V_{ij}\bar{u}_i
  \left[m_{u_i}P_L - m_{d_j}P_R\right]d_jH^+ 
  -\frac{\tan\beta}{v}m_{e_i}\bar{\nu}_i P_R e_iH^+
  + h.c.,
\end{equation}
As a result, such models are constrained by experimental bounds
on \(\mathrm{BR}(b\rightarrow s\gamma)\), \(\Delta M_K\), \(\Delta M_D\), 
\(\Delta M_B\),  rare Kaon decays, etc., which
translate into bounds on the model parameters relevant to the charged-scalar sector: 
$m_{H^\pm}$ and $\tan\beta$.    Since the flavor mixing in the charged Higgs couplings to the quark sector is proportional to $\cot\beta$, it is 
the region where both $\tan\beta$  and $m_{H^\pm}$  are small which is most tightly constrained by these bounds.
The most stringent constraints are those associated with \(b\rightarrow s\gamma\) and with
mixing in the \(B^{0}-\bar{B}^{0}\) and \(K_{L}-K_{S}\) systems.  In the 
L2HDM, the same Higgs doublet couples to both up- and down-type 
quarks, just as it does in Type~I 2HDM~\cite{Gunion:1989we,Haber:1978jt};
hence the bounds on $m_{H^\pm}$ and $\tan\beta$ due to flavor mixing in the quark sector will be essentially identical to those
applicable in Type~I models.  We now turn to review the bounds from each of these processes, updating the results obtained in~\cite{Barger:1989fj,Grossman:1994jb}.

\begin{figure}[t!]
  \begin{center}
\epsfxsize 2.3 truein \epsfbox {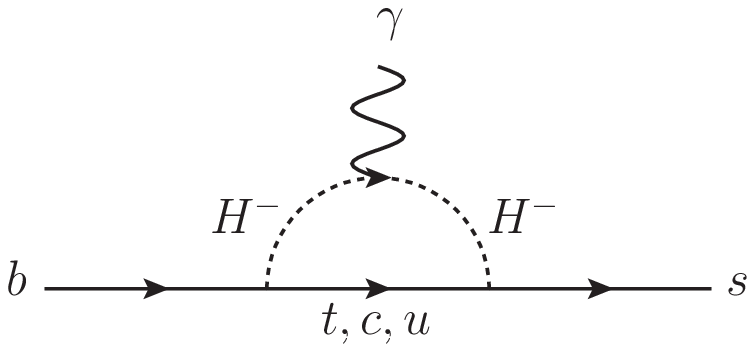} ~~~
\epsfxsize 2.3 truein \epsfbox {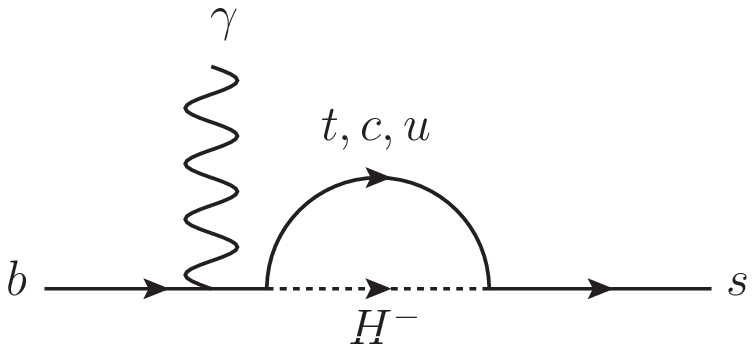} \\
\epsfxsize 2.3 truein \epsfbox {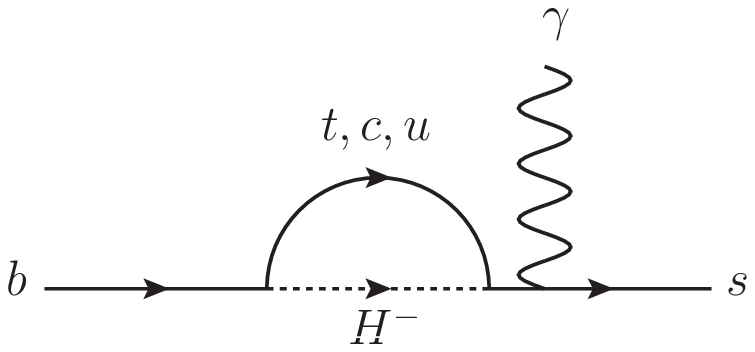} ~~~
\epsfxsize 2.3 truein \epsfbox {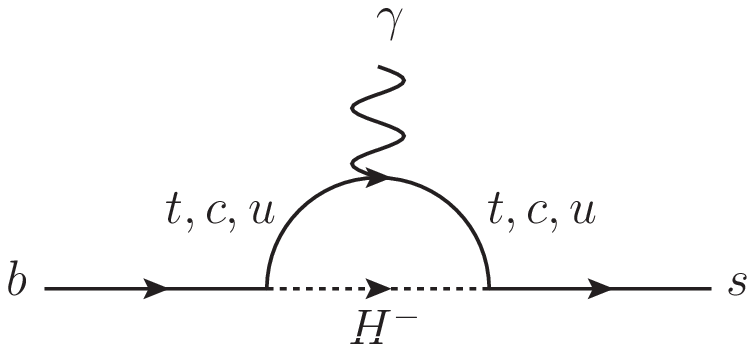} 
\end{center}
    \caption{The leading-order diagrams that yield a contribution to the
    \(b\rightarrow s\gamma\) amplitude due to the presence of massive,
    charged Higgs bosons in loops.  The Standard-Model contribution to
    this process involves similar diagrams with $H^-$ replaced by $W^-$.
    \label{fig:btosgammaDiagrams}}
\end{figure}

The first bounds we consider are those associated with the observed branching
ratio for the flavor-violating decay \(b\rightarrow s\gamma\).  
The combined
result from the CLEO and Belle experiments~\cite{Yao:2006px} is
\begin{equation}
  {\rm BR}(b\rightarrow s\gamma)=(3.3\pm0.4)\times 10^{-4}.
\end{equation}
This is consistent with the expected Standard Model result 
${\rm BR}^{SM}(b\rightarrow s\gamma)=3.32\times10^{-4}$.
In models with a non-minimal Higgs sector, additional contributions
to the amplitude for \(b\rightarrow s\gamma\) arise at the loop
level from diagrams involving virtual charged Higgs bosons, as discussed
above.  These
diagrams are compiled in Fig.~\ref{fig:btosgammaDiagrams} 
(SM contributions to this amplitude come from diagrams 
of the same sort,
but with \(W^{\pm}\) in place of \(H^{\pm}\).)  The rate for the
process can be calculated in the usual manner.  After incorporating
the effect of QCD corrections (which can be
quite large~\cite{Deshpande:1987nr}), one finds that~\cite{Barger:1989fj,Hou:1987kf}
\begin{equation}
  \Gamma(b\rightarrow
  s\gamma)=\frac{\alpha G_{F}^{2}m_{b}^{5}}{128\pi^{4}}|c_{7}(m_{b})|^2,
\end{equation}
where $c_7(m_b)$ is the coefficient of the effective operator
\begin{equation}
\mathcal{O}_7 \equiv F_{\mu\nu}\bar{s}_L\sigma^{\mu\nu}b_R
\end{equation}
in the conventions of Ref.~\cite{Grinstein:1987pu}, evaluated at the scale $m_b$.  
This coefficient takes the form 
\begin{eqnarray}
  c_{7}(m_{b})&=&\left(\frac{\alpha_{3}(M_{W})}{\alpha_{3}(m_{b})}\right)^{16/23}
  \times\nonumber\\ & &
  \left[c_{7}(M_{W})-\frac{3c}{10}\left[
  \left(\frac{\alpha_{3}(M_{W})}{\alpha_{3}(m_{b})}\right)^{10/23}-1\right]-
  \frac{3x}{28}\left[
  \left(\frac{\alpha_{3}(M_{W})}{\alpha_{3}(m_{b})}\right)^{28/23}-1\right]\right],
\end{eqnarray}
where the weak-scale amplitude function $c_7(M_W)$ in the L2HDM is given by 
\begin{equation}
  c_{7}(M_{W})=\sum_{i=u,c,t}V^{\ast}_{is}V_{ib}
  \left[G_{W}(x_{i})-\cot^{2}\beta G_{H}^{(1)}(y_{i})+\cot^{2}\beta
  G_{H}^{(2)}(y_{i})\right].
  \label{eq:c7weak}
\end{equation}
In these formulae, \(\alpha_{3}=g_{3}^{2}/4\pi\) and
\(\alpha=e^2/4\pi\), \(x_{i}=m_{q_{i}}^{2}/M_{W}^{2}\),
\(y_{i}=m_{q_{i}}^{2}/m_{H^{\pm}}^{2}\), \(G_{F}\) is the Fermi
constant, \(V_{ij}\) are elements in the CKM matrix, and \(c=232/81\). The functions \(G_{W}(x)\),
\(G_{H}^{(1)}(x)\), and \(G_{H}^{(2)}(x)\), which represent the loop
integral contributions to the \(b\rightarrow s\gamma\) amplitude, are
given in~\cite{Barger:1989fj}.

\begin{figure}[ht!]
  \begin{center}
  \epsfxsize 4 truein \epsfbox {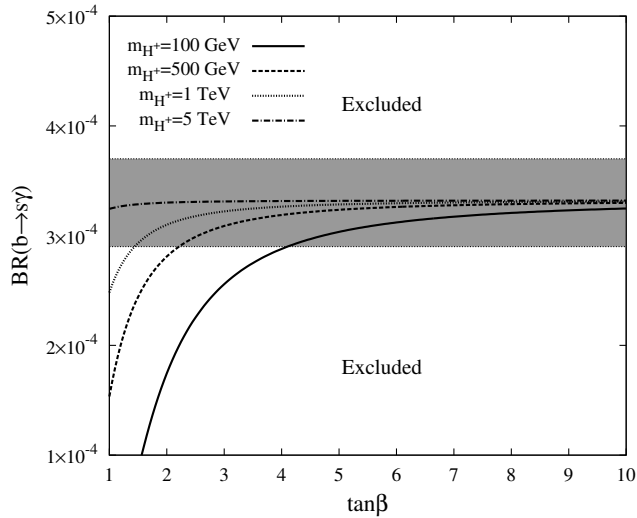}
  \end{center}
  \caption{Constraints on the charged Higgs mass and \(\tan\beta\)  from
  \({\rm BR}(b\rightarrow s\gamma)\) measurements.  The shaded horizontal band 
  corresponds to the experimentally-allowed $1\sigma$ region from CLEO and 
  Belle~\cite{Yao:2006px}.  The curves plotted here correspond 
  to $m_{H^\pm}=100$~GeV (solid line), $m_{H^\pm}=500$~GeV (dashed line), $m_{H^\pm}=1$~TeV (dotted line), and $m_{H^\pm}=5$~TeV (dash-dotted line).  
  \label{fig:bsgammaGraph}} 
\end{figure}

The constraints on \(m_{H^{\pm}}\) and \(\tan\beta\) from 
\(b\rightarrow s\gamma\) are
displayed in Fig.~\ref{fig:bsgammaGraph}.  Each curve therein 
represents the value of $\mathrm{BR}(b\rightarrow s\gamma)$ for
a given choice of $m_{H^\pm}$ as a function of $\tan\beta$.  Note that 
for the case under consideration here, in which 
$m_{H^\pm}>500\mbox{~GeV}$, the experimental constraints are
satisfied as long as $\tan\beta\gtrsim 2$. 

\begin{figure}[t!]
  \begin{center}
\epsfxsize 2.3 truein \epsfbox {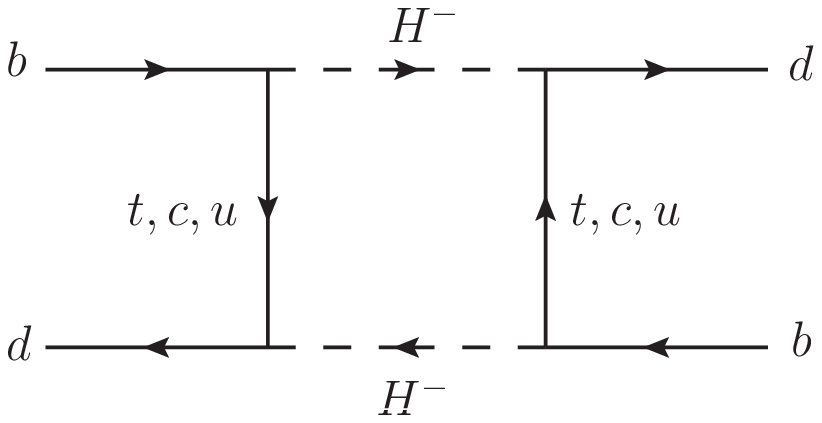} ~~~~~~~~~
\epsfxsize 2.3 truein \epsfbox {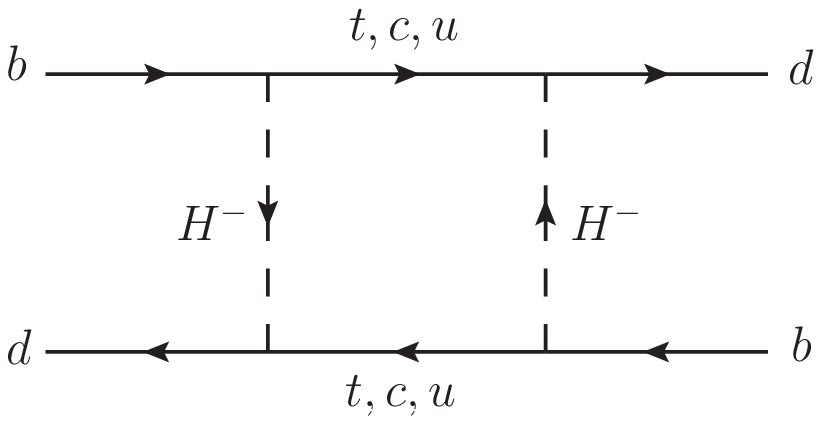} 
\end{center}
    \caption{Box diagrams that contribute to  $\Delta M_B$  in the \(B^{0}-\bar{B}^{0}\) 
    system via charged Higgs exchange.  Diagrams in which one or both of the 
    $H^-$ propagators is replaced by a $W^-$ propagator also contribute.  
    The box diagrams for transitions in the \(K_{L}-K_{S}\) system, etc.\ 
    are analogous. 
    \label{fig:MesonBoxDiagram}}
\end{figure}

Constraints on \(m_{H^{\pm}}\) and $\tan\beta$ can also be obtained from
limits on the observed mixing in the mesonic 
\(B^{0}-\bar{B}^{0}\) and \(K_{L}-K_{S}\) systems.  
The diagrammatic contributions to
\(B^{0}-\bar{B}^{0}\) mixing are shown in
Fig.~\ref{fig:MesonBoxDiagram}, and these contributions translate
into shift in the mass-splitting \(\Delta M_{B}\) between \(B^{0}\)
and \(\bar{B}^{0}\).  In the L2HDM, this splitting, including SM
contributions, is given by~\cite{Barger:1989fj}
\begin{eqnarray}
  \Delta M_{B}&=&\frac{G_{F}^{2}M_{W}^{2}}{6\pi^2}m_{B}f_{B}^{2}B_{B}
  \sum_{i=u,c,t}|(V_{ib}V_{id}^{\ast})^{2}|\eta_{\mathit{QCD}}
  \nonumber\\ & &\left[A_{WW}(x_{t})+\cot^{4}\beta
  A_{HH}(x_{t},x_{H},x_{b})+\cot^{2}\beta A_{WH}(x_{t},x_{H},x_{b})
  \right],\label{eq:DeltaMBB}
\end{eqnarray}
where the \(x_{i}\) are defined as below Eqn.~(\ref{eq:c7weak}),
\(f_{B}\) is the \(B\)-meson decay constant, \(B_{B}\) is the ``bag
factor'' (which encompasses all deviations from the vacuum
saturation approximation).  Expressions for the factor 
\(\eta_{\mathit{QCD}}\), which
accounts for QCD effects, along with the functions \(A_{WW}(x_{t})\),
\(A_{HH}(x_{t},x_{H},x_{b})\), and \(A_{WH}(x_{t},x_{H},x_{b})\)
can be found in~\cite{Barger:1989fj}.

As for \(f_{B}\) and \(B_{B}\), there is a good deal of uncertainty
as to their precise numerical values.  Since they appear in
Eqn.~(\ref{eq:DeltaMBB}) in the combination \(f_{B}B_{B}^{1/2}\),
it is easier simply to deal with the uncertainty in this single
quantity.  Estimates of \(f_{B}B_{B}^{1/2}\) have been made using a
variety of  lattice QCD sum rules in conjunction with
experimental evidence on heavy meson decays from SLAC, and the
uncertainties in their values depend on the summation methods
employed and the assumptions made.
Following~\cite{Barger:1989fj, Franzini:1988fs}, we take the range of
uncertainty to be
\begin{equation}
  100\mathrm{~MeV}\lesssim f_{B}B_{B}^{1/2}\lesssim
  180\mathrm{~MeV}.
  \label{eq:uncertaintyfBrootB0}
\end{equation}

Instead of dealing with \(\Delta M_{B}\) directly, it is
easier to deal with the combination
$x_{d}\equiv\Delta M_{B}/\Gamma_{B}$,
since the time-integrated mixing probability in the
\(B^{0}-\bar{B}^{0}\) system depends on this combination of
variables.  The accepted experimental value for \(x_{d}\), as reported 
by the Heavy Flavor Averaging Group, is $x_{d}=0.776\pm0.008$~\cite{Yao:2006px}.  
Using the observed lifetime of the \(B^{0}\) meson
($\tau_{B}=1.530\times 10^{-12}\mathrm{~sec}$)
and the expression in Eqn.~(\ref{eq:DeltaMBB}),  one may obtains a 
theoretical value for \(x_{d}\), which can be compared to
this experimental result.  

\begin{figure}[ht!]
  \begin{center}
    \epsfxsize 3.75 truein \epsfbox {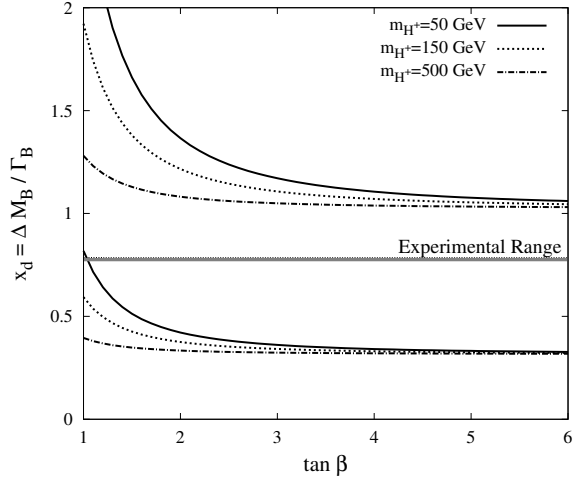}
  \end{center}
    \caption{Bounds on $m_{H^\pm}$ and \(\tan\beta\) from mixing in the
    \(B^{0}-\bar{B}^{0}\) system, plotted as a function of $\tan\beta$.
    The thin shaded region represents the experimentally-allowed $1\sigma$ 
    range for $x_d=\Delta M_B/\Gamma_B$~\cite{Yao:2006px}.    
    Each pair of thick curves represents the upper and lower limits on 
    the theoretical value of $x_d$ (due to uncertainties in hadronic 
    matrix elements, etc.) for three
    different choices of $m_{H^\pm}$: $50$~GeV (solid lines), 
    $150$~GeV (dotted lines), and $500$~GeV (dot-dashed lines).  A certain
    combination of $\tan\beta$ and $m_{H^\pm}$ is permitted as long as any part of the
    experimentally-allowed range falls between the lines corresponding to the 
    upper and lower theoretical limits. 
      \label{fig:MixingGraphs}}
\end{figure}

In  Fig.~\ref{fig:MixingGraphs}, we
show how the \(B^{0}-\bar{B}^{0}\) mixing bound constrains
\(m_{H^{\pm}}\) and \(\tan\beta\). As there is a large uncertainty
in \(f_{B}B_{B}^{1/2}\)~[Eqn.~(\ref{eq:uncertaintyfBrootB0})], and in fact
one far larger than that associated with the measured value of
\(x_{d}\), the theoretical prediction for a given choice of
 \(m_{H^{\pm}}\) translates into a broad band in 
\(\tan\beta-x_{d}\) space, rather than a thin line.  In
Fig.~\ref{fig:MixingGraphs}, the upper and lower bounds of each such 
band are demarcated by a pair of thick, dark lines of the same type 
(solid, dotted and dot-dashed). The thin, shaded, horizontal stripe 
represents the experimentally-allowed window.  If
any part of this stripe falls within the band corresponding to a given
value of  $m_{H^\pm}$ for a given $\tan\beta$, that parameter combination
is permitted by the $\Delta M_B$ constraint.  We see from the plot that
this constraint only becomes relevant for very small values of 
$\tan\beta\sim 1$, and thus is not particularly stringent.   

Similar calculations to those outlined above for the
\(B^{0}-\bar{B}^{0}\) system can also be performed for mixing
in the \(K_{L}-K_{S}\) and \(D^{0}-\bar{D}^{0}\) 
systems~\cite{Barger:1989fj}.  In addition, limits
can also be derived on the $CP$-violating parameters $\epsilon$ and
${\epsilon}'$.  However, due to large theoretical uncertainties in the 
hadronic matrix elements, the resulting bounds on new physics from
these considerations are not particularly stringent in the L2HDM,
especially when $\tan\beta>1$~\cite{Grossman:1994jb}.

Experimental bounds on leptonic charged-meson decays --- 
$D_{S}^{\pm}\rightarrow\mu^\pm\nu$, 
$D_{S}^{\pm}\rightarrow\tau^\pm\nu$, 
$K^{\pm}\rightarrow\mu^\pm\nu$,
$B^{\pm}\rightarrow\tau^\pm\nu$ etc.\ --- can also
be used to constrain 2HDM~\cite{Hou:1992sy}.
In general, the partial width for the leptonic decay of a given meson is modified by a $\tan\beta$-dependent factor $r_{M\ell}$, which
in many scenarios (e.g.\ Type~II models) can be quite sizeable when $\tan\beta$
is large \cite{Akeroyd:2009tn}.   In the L2HDM, however, the
$r_{M\ell}$ are independent of $\tan\beta$ due to the cancellation of the $\tan\beta$ factors between the quark and the lepton couplings.  As a result the model is essentially
unconstrained by these considerations.  Experimental limits on the rates for leptonic 
decays such as $\tau\rightarrow\mu\bar{\nu}\nu$ can also constrain models with enhanced
Higgs couplings to leptons~\cite{Krawczyk:2004na}.  However, such constraints
only become relevant when the charged-Higgs mass is $\mathcal{O}(100\mbox{~GeV})$ or lower,
or when $\tan\beta$ is extremely large, and thus have little bearing on the 
decoupling regime studied here. 

The above analysis shows that in the decoupling region (as we have defined it), 
where $m_{H^\pm}>500$~GeV, all constraints from direct charged-Higgs searches, 
neutral meson mixing, $CP$-violation, charged-meson decay, etc.\ can be satisfied as 
long as \(\tan\beta\) is greater than $\sim 2$.
This is mainly due to the fact that in the L2HDM, there is no new source 
of flavor violation except the SM CKM matrix.  The effective couplings between $H^\pm$ 
and the SM quarks are proportional to $\cot\beta$, which implies that the most 
stringent constraints become weaker as $\tan\beta$ increases.
Thus, we conclude that experimental constraints from
flavor violation, direct searches, etc.\ do not
pose any significant issues for the L2HDM as long as the charged Higgs scalars are heavy.
(Indeed, a relatively low value of
\(\tan\beta\approx 3\) and a charged Higgs light enough
to be discovered at the LHC are by no means incompatible.)
This is true even in the region of parameter space most interesting 
for collider physics, in which both $\sin\alpha$ and $\tan\beta$ are large, and the effective couplings between the lightest
$CP$-even Higgs and the SM leptons differ drastically from their SM values.

\section{Branching Ratios and Cross-Sections\label{sec:BRXSec}}

We now turn to examine the effect of these coupling modifications on the
production cross-sections and decay widths of a light Higgs boson.
Since the overall amplitudes for Higgs decays into any two-particle final state
\(X\) scale as \(|\eta_{X}|^2\) (i.e., the appropriate \(\eta\)-factor for that final
state), the associated branching ratios scale like
\begin{equation}
  \frac{\mathrm{BR}(h\rightarrow X)}{\mathrm{BR}^{SM}(h\rightarrow X)}=
    |\eta_{X}|^2
  \frac{\Gamma_{\mathit{tot}}^{SM}(h)}{\Gamma_{\mathit{tot}}(h)}
=
  |\eta_{X}|^2\left(\sum_i |\eta_{Y_i}|^2 \mathrm{BR}^{SM}(h\rightarrow Y_i)\right)^{-1}.
\end{equation}

\begin{figure}[ht!]
  \begin{center}
    \epsfxsize 2.5 truein \epsfbox{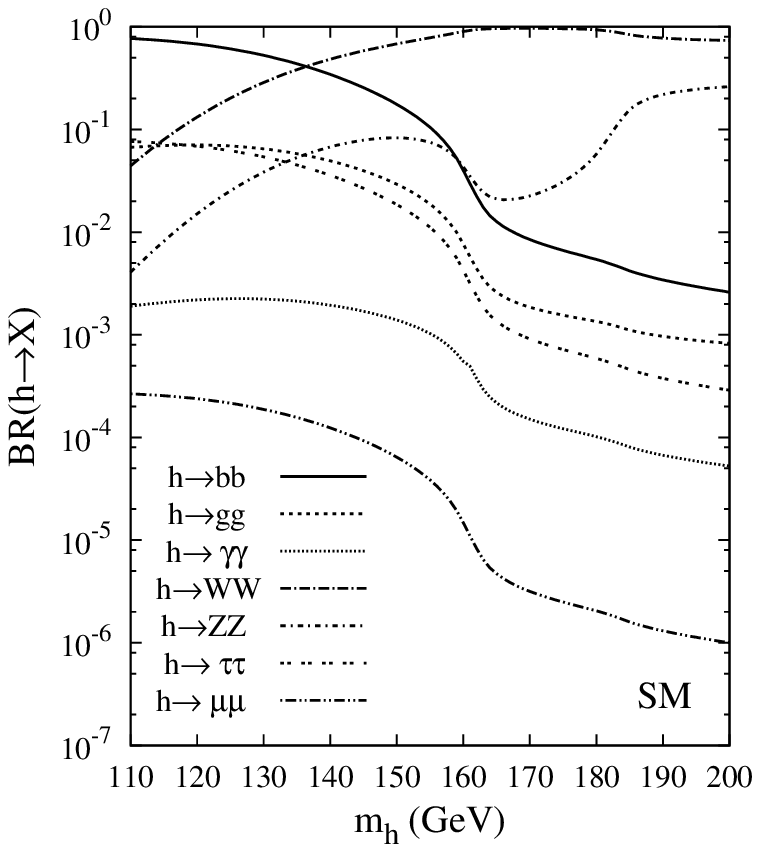}
    \epsfxsize 2.5 truein \epsfbox{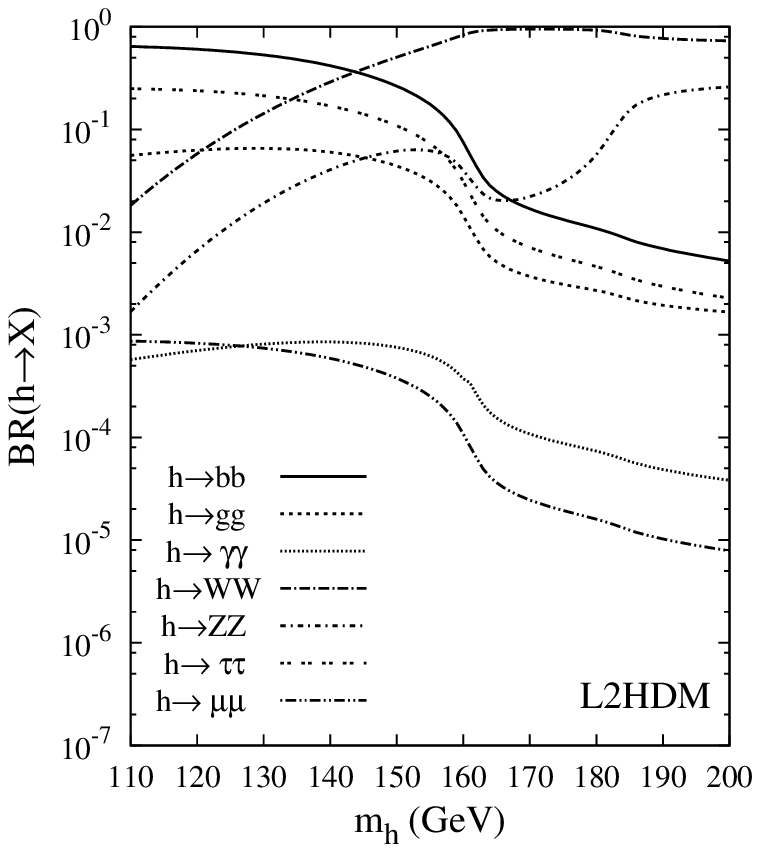}
  \end{center}
    \caption{Plots of the Branching ratios for the a number of two-body decays of the 
    Higgs boson, both in the SM (left panel) and in the
   L2HDM (right panel) for the benchmark point \(( \sin\alpha=0.55,\ \tan\beta=3) \).  Note that $\mathrm{BR}(h\rightarrow\tau^+\tau^-)$ 
    and $\mathrm{BR}(h\rightarrow b\bar{b})$ are on the same order, as are
    $\mathrm{BR}(h\rightarrow\mu^+\mu^-)$ and $\mathrm{BR}(h\rightarrow \gamma\gamma)$. 
  \label{fig:BRPlot}}
\end{figure}

In order to provide a concrete example of the effect such a modification can have on
Higgs phenomenology, let us focus on a particular benchmark point: 
\(\sin\alpha=0.55\), \(\tan\beta=3\),
which we have indicated by a dot in Fig.~\ref{fig:SinaTanbExcl}. 
We pick this particular point as a benchmark because it yields only a moderate  
deviation from the SM couplings and is consistent with the 
bounds~(\ref{eq:PertBounds}) and~(\ref{eq:VacStabBounds}) when
$m_{H^\pm},m_H,m_A > 500\mbox{~GeV}$.  The $\eta$-factors corresponding to this 
particular point are  
\begin{equation}
\eta_q=\eta_g=0.88,~~~ \eta_{\ell}= -1.74,~~~ \eta_V=0.62,~~~ \eta_\gamma=0.54.
\label{eq:EtasforBM}
\end{equation}
Fig.~\ref{fig:BRPlot} illustrates the effect of this coupling-constant modification
on the branching ratios of a light, $CP$-even Higgs scalar.  In the left-hand panel,
we have plotted the SM branching ratios for a number of Higgs decay processes 
as a function of $m_h$.
All branching ratios used in the construction of the figure 
were calculated using HDECAY~\cite{Djouadi:1997yw}.
In the right-hand panel, we have plotted 
branching ratios for the same set of processes in the L2HDM at our chosen benchmark point.  It is evident that
even this moderate modification of the couplings has a dramatic effect on
the decay behavior of a light Higgs: for example, the rates for 
$\mathrm{BR}(h\rightarrow\tau^+\tau^-)$ and 
$\mathrm{BR}(h\rightarrow b\bar{b})$ are on the same order.  Since
$h\rightarrow b\bar{b}$ is the dominant decay channel for a Higgs boson
with a mass in the range $114\mbox{~GeV}\lesssim m_h\lesssim 140\mbox{~GeV}$,
this clearly represents a substantial effect on Higgs phenomenology.  It is
also worth noting that $\mathrm{BR}(h\rightarrow \mu^+\mu^-)$ and
$\mathrm{BR}(h\rightarrow \gamma\gamma)$ are also on the same order for this
choice of parameters.  This suggests that processes involving
direct decays of a light Higgs boson to a pair of high-$p_T$ muons
could play an important role in the collider phenomenology of the light Higgs $-$ a suggestion
we will explore further in Section~\ref{sec:LHCSignatures}.  
The branching ratios for a number of other decay channels relevant to 
the study of a light SM Higgs boson at the
LHC, such as $h\rightarrow  \gamma\gamma$, are clearly 
suppressed here relative to their SM values.  
          
\begin{figure}[ht!]
  \begin{center}
    \epsfxsize 4.0 truein \epsfbox{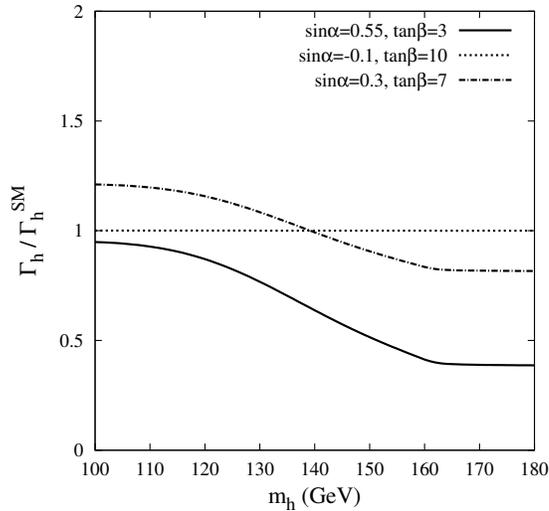}
  \end{center}
    \caption{Plot of the ratio of the total width of the Higgs boson in the leptophilic
    2HDM to that of a SM Higgs for a representative sample of points in $\sin\alpha$ -
    $\tan\beta$ parameter space as a function of the Higgs mass $m_h$.
  \label{fig:GamTotPlot}}
\end{figure}

The effect of the coupling modifications on the total Higgs width is shown in Fig.~\ref{fig:GamTotPlot}.  Here, we have plotted the ratio of the total Higgs width
$\Gamma_{\mathit{tot}}(h)$ to its SM value for three different points in the allowed
region of $\sin\alpha$ $-$  $\tan\beta$ parameter space as a function of $m_h$.  The first
of these points is our chosen benchmark ($\sin\alpha=0.55$, $\tan\beta=3$), for which
$\Gamma_{\mathit{tot}}(h)$ (indicated by the solid line) is slightly lower than 
its SM value due to the suppression of $\Gamma(h\rightarrow b\bar{b})$ when $m_h$ is small.
When $m_h$ increases and decays to electroweak gauge bosons begin to dominate the Higgs width,
$\Gamma_{\mathit{tot}}(h)$ drops even further, since $\eta_V < \eta_q$ at this point~[see Eqn.~(\ref{eq:EtasforBM})].
The second of these points, ($\sin\alpha=-0.1$, $\tan\beta=10$), is located very near the ``pure decoupling"
line in Fig.~\ref{fig:SinaTanbExcl}; hence
for this point $\Gamma_{\mathit{tot}}(h)$ (indicated by the dotted line) is essentially
equal to $\Gamma_{\mathit{tot}}^{SM}(h)$.  At the third point, 
($\sin\alpha=0.3$, $\tan\beta=7$), a substantial enhancement in 
$\Gamma(h\rightarrow \tau\tau)$ overcomes the suppression factor in
$\Gamma(h\rightarrow b\bar{b})$, and consequently 
$\Gamma_{\mathit{tot}}(h)>\Gamma_{\mathit{tot}}^{SM}(h)$ for $m_h\lesssim 140$~GeV
(as indicated by the dash-dotted line).  For
larger values of $m_h$, gauge-boson decays once again dominate the Higgs 
width, which becomes suppressed relative to its SM value.       
Even in the most extreme cases permitted by 
the model consistency constraints outlined in Section~\ref{sec:Param}, however,
$\Gamma_{\mathit{tot}}(h)/\Gamma_{\mathit{tot}}^{SM}(h)\lesssim 2$.  This implies
that the narrow-width approximation remains valid over the Higgs mass range 
$114\mbox{~GeV}\lesssim m_h\lesssim 140\mbox{~GeV}$, which will be the mass region of primary focus
of the present work.

Since we have shown that the narrow-width approximation to be valid, we can proceed
in a straightforward manner from the decay width calculations above to determine
how the  
cross-sections for full collider processes are modified.  In this approximation,
one assumes that essentially all the Higgs bosons produced in any such process are 
produced on-shell.  This allows one to approximate the cross-section for any
process of the form \(Y\rightarrow h \rightarrow X\) by
\begin{equation}
  \sigma(Y\rightarrow h \rightarrow X)\approx \sigma(Y\rightarrow h)\times\mathrm{BR}
  (h\rightarrow X).
\end{equation}
Furthermore, if the SM production cross-section \(\sigma^{\mathit{SM}}(Y\rightarrow h)\)
for the process is known, one can use the fact that \(\sigma(Y\rightarrow h)\propto
\mathit{\Gamma}(h\rightarrow Y)\) to obtain the relation
\begin{eqnarray}
  \frac{\sigma(Y\rightarrow h \rightarrow X)}{\sigma^{\mathit{SM}}(Y\rightarrow h \rightarrow
  X)}&=&\frac{\Gamma(h\rightarrow Y)}{\Gamma^{\mathit{SM}}(h\rightarrow Y)}
  \times
  \frac{\mathrm{BR}(h\rightarrow X)}{\mathrm{BR}^{\mathit{SM}}(h\rightarrow X)}\nonumber\\&=&
  \frac{\mathrm{BR}(h\rightarrow Y)}{\mathrm{BR}^{\mathit{SM}}(h\rightarrow Y)}\times
  \frac{\mathrm{BR}(h\rightarrow X)}{\mathrm{BR}^{\mathit{SM}}(h\rightarrow X)}\times
  \frac{\Gamma_{\mathit{tot}}(h)}{\Gamma_{\mathit{tot}}^{\mathit{SM}}(h)},
  \label{eq:ExSecRats}
\end{eqnarray}
which allows us to calculate the cross-sections for these overall processes in the
modified model.     

For the benchmark point that we have chosen ($\sin\alpha=0.55$, $\tan\beta$=3), the 
cross-sections for most of the conventional Higgs search modes at the LHC are 
suppressed relative to their SM values, due to the suppressed Higgs couplings to quarks and to gauge bosons. 
Many of the processes in which the Higgs decays directly to 
charged-lepton pairs, on the other hand, 
are substantially enhanced.  We will discuss the implications these 
modifications can have for Higgs
searches in detail in Section~\ref{sec:Combined}.

\section{LHC Signatures of a Leptophilic Higgs Boson
\label{sec:LHCSignatures}}

One of the most interesting aspects of the L2HDM is that in 
certain regions of parameter space, new channels for the discovery of a light Higgs 
boson can open up.  In particular, when the effective coupling between $h$ 
and the SM leptons is substantially increased while its 
couplings to SM quarks and/or electroweak gauge bosons are not dramatically 
suppressed, a number of processes in which the Higgs boson decays directly to a 
pair of high-$p_T$ leptons can become far more important for the discovery of a 
light Higgs than they are in the SM. 
In our analyses, we focus on the discovery of $h$ in the light-to-intermediate-mass 
region $120\ {\rm GeV}<m_h<140\ {\rm GeV}$.  For heavier Higgs bosons, 
$h\rightarrow WW^*,ZZ^*$ dominates and leptonic Higgs decays play a 
less important role.
In the decoupling limit case studied here,
in which the additional Higgs scalars $H^{\pm}$, $H$, and $A$ are heavy, such 
processes might well constitute the only evidence for physics beyond the Standard 
Model accessible within the first $30\mbox{~fb}^{-1}$ of integrated luminosity at the LHC,  
and are therefore of crucial importance.  This situation is quite different from the 
one studied in Ref.~\cite{Goh:2009wg}, in which some of these additional
scalars are light and play a significant role 
in collider phenomenology.

Since the largest leptonic contribution to the Higgs total width comes from
$h\rightarrow\tau\tau$, processes involving Higgs decays directly to tau leptons will
play a significant role in the collider phenomenology of the L2HDM.  However,
the analysis of such processes is complicated by subtleties associated with tau 
decay.  Each \(\tau\) lepton can decay either leptonically or hadronically,  
with respective branching ratios~\cite{Yao:2006px}, 
${\rm BR}_\tau^{\mathit{lep}}\simeq 35.20\%$ and
${\rm BR}_\tau^{\mathit{had}}\simeq 64.80\%$.
We will henceforth denote a hadronically-decaying tau as \(\tau_h\) and a leptonically
decaying one as \(\tau_{\ell}\).  For processes involving \(CP\)-even Higgs boson 
decays into a \(\tau\tau\) pair, there are two final states which permit
successful identification of both taus: $e\mu + \displaystyle{\not} E_T$ and
$\tau_h\ell+\displaystyle{\not} E_T$, where \(\ell=e,\mu\).  Final states resulting from
fully hadronic decays have a large background from dijet processes 
with narrow jets misidentified as taus.  Final states involving two leptons 
of like flavor (\(e^+e^-+\displaystyle{\not} E_T\) 
and \(\mu^+\mu^-+\displaystyle{\not} E_T\)) are also less useful due to the overwhelming SM background from $Z/\gamma^\ast\rightarrow \ell^+\ell^-$ processes.

A hadronically-decaying tau will decay  
into either a ``one-prong'' (approximately 77\% of the time)
or ``three-prong'' (approximately 23\% of the time) final state.   
These final states involve narrow,
well-collimated jets including one or three charged pions, 
respectively.  The identification of a jet as coming from a 
hadronically-decaying \(\tau\), as opposed to some QCD process, is far from 
trivial.  One of the principal discrimination variables is jet radius \(R_{\mathit{EM}}\) 
(see~\cite{tauID} for more details regarding $\tau$ identification).  
At the Tevatron (Run II), \(\epsilon_{\tau_h}\approx 35\%-40\%\) for a \(p_T^\tau>20\)~GeV cut.
At the LHC, a $\tau$ identification efficiency of around \(50\%-60\%\) is 
expected~\cite{tauID}.

Processes involving direct decays of $h$ to muon pairs can also be of interest
for Higgs discovery in the L2HDM.  The disadvantage of
such channels for Higgs searches, relative to those involving direct decays to taus, 
is the suppressed branching ratio.  Since Yukawa coupling universality dictates that 
\(y_\mu/y_\tau\propto m_\mu/m_\tau\), both in the SM and in the L2HDM, \(\mathrm{BR}(h\rightarrow \mu\mu)\ll \mathrm{BR}(h\rightarrow \tau\tau)\).
However, this is compensated for to a great extent by the fact that the
dimuon signal is exceptionally clean.  Indeed, the muon identification efficiency 
at the LHC is more that 90\%~\cite{ATLASTDR, CMSTDR}.  In addition, the measurement of 
muon momenta allows for a precise reconstruction of the Higgs mass within 
$\pm 2.5$~GeV.  This permits the implementation of an extremely efficient cut on 
$M_{\mu\mu}$, the invariant  mass of the muon pair, and a substantial reduction 
in background levels for all channels involving direct Higgs-boson decays to
muon pairs.   

We now turn to address the prospects for detecting a light SM-like $CP$-even Higgs boson 
at the LHC on a channel-by-channel basis.
In the present work, as discussed in Section~\ref{sec:Param}, we will assume 
generation universality among the lepton Yukawa couplings.
Therefore, we will ignore the $h\rightarrow e e$ channel and focus only on $h\rightarrow \tau\tau$ and $h\rightarrow \mu\mu$. 
The channels of primary interest, then, are
those in which the Higgs is produced by gluon fusion, weak-boson fusion, 
or $t\bar{t}h$ associated production and decays to either $\mu^+\mu^-$ or 
$\tau^+\tau^-$.  Associated $W^\pm$ and $Z$
production processes generally have smaller rates, but may also potentially be of
interest, and as such we briefly discuss them as well.  Bottom-quark-fusion 
processes with a leptonically-decaying Higgs boson~\cite{Barger:1997pp}, 
while potentially interesting for Type~II 2HDM in which the 
$h\bar{b}b$ vertex receives a large $\tan\beta$-enhancement, are less useful in the
L2HDM, since the effective down-type quark couplings are
suppressed in that scenario rather than enhanced.
In this section, we briefly summarize the results of the existing studies of the 
leptonic-Higgs-decay channels at the LHC, with an eye toward their utility for
the discovery of a leptophilic Higgs.

\subsection{$qq'\rightarrow qq'h(h\rightarrow \tau\tau)$}

We begin with a discussion of the weak-boson-fusion process 
$qq'\rightarrow qq'h(h\rightarrow \tau\tau)$, which is the only channel involving direct Higgs-boson decay to a pair of charged leptons that
contributes significantly to the Higgs discovery potential in the 
SM.  Indeed, it is a particularly promising channel for SM Higgs discovery in 
the intermediate mass region ($125\mbox{~GeV}\lesssim m_h\lesssim 140\mbox{~GeV}$)~\cite{Rainwater:1998kj}.  Discriminating between signal and SM background    
can be facilitated by requiring that events have two leading tagging jets in the 
forward-backward direction and imposing a minijet veto in the central region of the detector.  
A great deal of attention has been devoted to this channel, with an emphasis on 
$\tau_h\tau_\ell$ and $\tau_\ell\tau_\ell$ final states.  Combining all channels, a 
statistical significance of more than $5\sigma$ can be reached for Higgs masses around 
120 $-$ 130 GeV with $30\ {\rm fb}^{-1}$ of integrated luminosity at ATLAS~\cite{Asai:2004ws}.  
The detection prospects are similar at CMS~\cite{Abdullin:2005yn}.
 
\subsection{$gg\rightarrow h\rightarrow \tau\tau$}

The prospects for detecting a light, SM Higgs boson produced
by gluon fusion and decaying to \(\tau^+\tau^-\) at the Tevatron were
examined in Ref.~\cite{Belyaev:2002zz}.  
In order to effectively reconstruct the Higgs mass from the various final-state 
particles produced during tau decay, it is necessary to focus on events in which 
the transverse momentum of the tau pair is nonzero; hence the
authors elected to focus on the process 
$p\bar{p}\rightarrow hj\rightarrow \tau^+\tau^-j$.
Taking into account both the $S/B$ and $S/\sqrt{B}$ ratios, $\tau_h\tau_\ell$ turns out 
to be the most promising channel for signal identification, but that an integrated 
luminosity of $14\mbox{~fb}^{-1}$ would be needed at the Tevatron in order to exclude 
a 120 GeV SM Higgs boson at the 95\% C.L.  However, preliminary 
studies at ATLAS~\cite{RichterWasNote} indicate that this will be a promising channel  in which to look for a Higgs boson with enhanced coupling to leptons at the LHC.

\subsection{$tth(h\rightarrow \tau\tau)$}

This process was examined in a Standard Model context in~\cite{Belyaev:2002ua}.
In order to be able to reconstruct the two top quarks effectively, the authors 
restricted their analysis to cases in which one of the \(W\) bosons produced 
during top decay decays leptonically, while the other decays hadronically.  Only events with 
hadronic tau decays were considered, as reconstructing both tops proves to be slightly 
easier in this scenario.  Thus the overall process of interest
is \(pp\rightarrow t\bar{t}h\rightarrow bbjj\ell {\tau}_h{\tau}_h+ 
\displaystyle{\not} E_T\). 
Since the production cross section drops quickly with increased Higgs mass, this channel is only important when the Higgs is light.  For $m_h$ around 120~GeV, a statistical significance of 
$4\sigma$ can be obtained with $100\mbox{~fb}^{-1}$ of integrated luminosity. 
In~\cite{Gross:2005tth}, semileptonic tau decays were considered --- in particular, decays 
of the form \(pp\rightarrow t\bar{t}h\rightarrow bbjj\ell{\tau}_h\tau_\ell + 
\displaystyle{\not} E_T\), and it was found that such a process could provide evidence for a
120~GeV Higgs boson at the $2.7\sigma$ level for an integrated luminosity of 
$30\mbox{~fb}^{-1}$.

\subsection{$qq'\rightarrow qq'h(h\rightarrow \mu\mu)$}

The weak boson fusion process \(qq'\rightarrow qq'h(h\rightarrow\mu\mu)\) 
was analyzed in~\cite{Plehn:2001qg}.   
After the appropriate cuts on the tagging jets are imposed, 
the leading SM background comes from irreducible 
$Zjj$ or $\gamma^\ast jj$ processes, with the $Z/\gamma^\ast$ decaying to muon pairs.  
Due to the extremely suppressed SM $h\rightarrow\mu\mu$ branching ratio, an integrated 
luminosity of \(\mathcal{O}(300\mbox{ fb}^{-1})\) or more is generally required to 
claim a $3\sigma$ discovery for a Higgs mass less than 140~GeV.  In the L2HDM, however the detection prospects can be substantially improved if 
$\eta_{V}\sim 1$ and \(\eta_{\ell}\gg 1\).

\subsection{$gg\rightarrow h\rightarrow \mu\mu$}

The prospects for the detection of a light Higgs boson of $110\mbox{~GeV}\leq m_{h}\leq 
140\mbox{~GeV}$ produced by gluon fusion and decaying directly into $\mu^+\mu^-$
were discussed in~\cite{Han:2002gp}.  The irreducible background for 
$gg\rightarrow h\rightarrow \mu\mu$ is dominated by the Drell-Yan processes
\(q\bar{q}\rightarrow Z^*/\gamma^*\rightarrow \mu\mu\).  The sharp invariant mass 
resolution of the muon pair allows for a substantial reduction in this background via
a stringent cut on $M_{\mu\mu}$.  Consequently, a significance level similar to 
that in the WBF channel as discussed in ~\cite{Plehn:2001qg} can be attained in this channel as well.

\subsection{$tth(h\rightarrow \mu\mu)$}

The prospects for discovering a Standard Model Higgs boson in the 
$tth(h\rightarrow \mu\mu)$ channel were recently studied in~\cite{Su:2008bj}.
This channel tends to be more important when the Higgs mass is light 
(around 120~GeV) since the production cross section drops 
quickly for a heavier Higgs. 
The primary irreducible backgrounds, which come from $t\bar{t}Z$ and $t\bar{t}\gamma^\ast$ 
production, with the $Z/\gamma^\ast$
decaying into a muon pair, can be reduced quite effectively by a cut on that
muon pair's invariant mass.  Additional, reducible backgrounds such as 
\(Zb\bar{b}jjjj\) can be effectively eliminated by reconstructing the
masses of both top quarks, which is possible in the case where the tops decay 
either fully hadronically or semileptonically.

The statistical significances for the $tth(h\rightarrow \mu\mu)$ channel
are  of roughly the same
order as those in the $gg\rightarrow h\rightarrow \mu\mu$ and 
$qq'\rightarrow qq'h(h\rightarrow\mu\mu)$ channels, and hence could contribute
significantly to the discovery potential for a light Higgs scalar with enhanced 
couplings to leptons. 

\subsection{$Wh/Zh(h\rightarrow \tau^+\tau^-)$ and $Wh/Zh(h\rightarrow \mu^+\mu^-)$}

Higgs production via the processes \(pp\rightarrow Wh\) and \(pp\rightarrow Zh\) could
also potentially play a role in the discovery of a leptophilic Higgs, though the prospects
in these channels are not as favorable as the other, aforementioned ones.  SM cross-sections
for these processes, taking into account the leptonic decay of the Higgs boson, are given  
in Table~\ref{table:WHZH} for the case in which $m_h=120$~GeV.  These were 
determined from leading-order results obtained using MADGRAPH~\cite{Alwall:2007st} 
and modified by the appropriate $K$-factors: $K_S=1.27$ for signal~\cite{Brein:2003wg}, 
$K_{BG}=1.7$ for background~\cite{Frixione:1992pj}.  
For processes in which the Higgs decays to $\mu^+\mu^-$, the signal is clearly too 
small to be of any use.  However, for processes involving decays to $\tau^+\tau^-$, 
the signal is only about a factor of $\sim 25$ smaller than the background. 
By optimizing 
cuts to eliminate the SM background, this channel might potentially be 
of use --- particularly if $\mathrm{BR}(h\rightarrow \tau\tau)$ is enhanced, 
as in the L2HDM.  Little analysis of these processes exists in the literature, 
and we leave the detailed study of these channels for future work.

\begin{table}
\begin{center}
\begin{tabular}{|c|c|c|c|c|c|}\hline
&Signal (fb)&BG (fb)&&Signal(fb)&BG(fb) \\\hline
$Zh(h\rightarrow \mu\mu)$&$0.113$  &$1156.5$&$Zh(h\rightarrow \tau\tau)$&$32.58$  &$1156.5$\\
$Wh(h\rightarrow \mu\mu)$ &$0.215$  &$1534.3$&$Wh(h\rightarrow \tau\tau)$&$61.85$ &$1534.3$ \\\hline
\end{tabular}
\caption{SM production cross sections at the LHC for $Wh$, $Zh$ associated production, with $h$ decays into muon pair or tau pair.  The Higgs mass is taken to be 120 GeV. Also shown are the SM background  $ZZ$, $WZ$ with one $Z$ decays into muons or taus.    The numbers are obtained using MADGRAPH~\cite{Alwall:2007st}.
}
\label{table:WHZH}
\end{center}
\end{table}
  
\section{LHC Discovery Potential\label{sec:Combined}}

\begin{figure}[ht!]
  \begin{center}
    \epsfxsize 2.75 truein \epsfbox{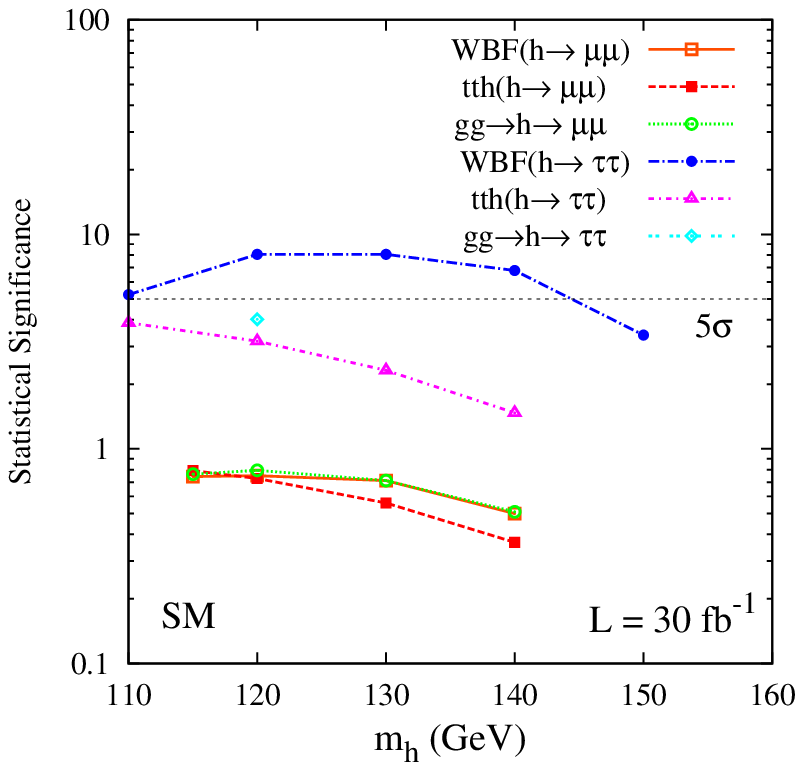}~~~~
    \epsfxsize 2.75 truein \epsfbox{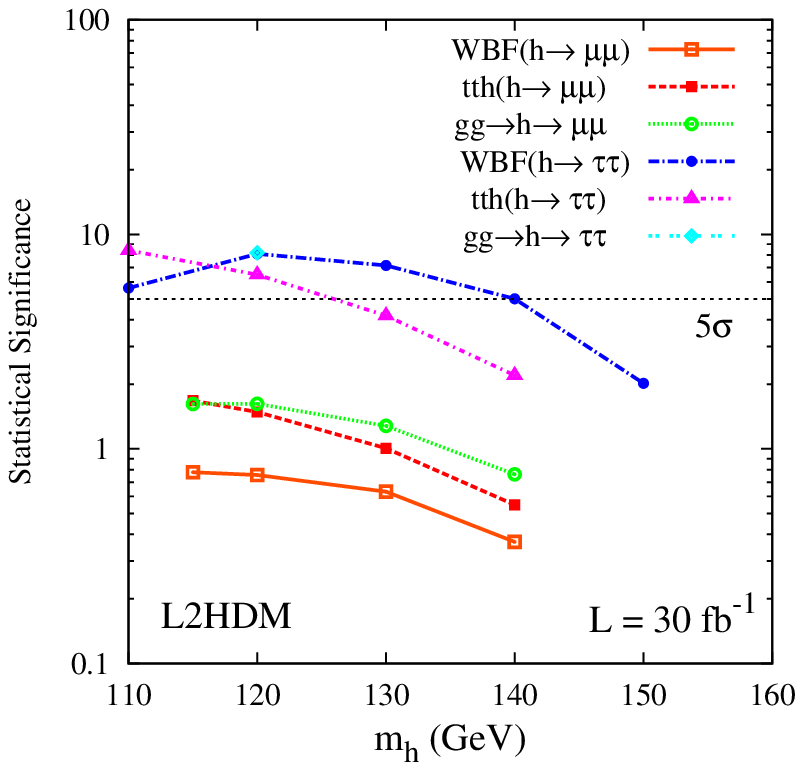}
  \end{center}
    \caption{Plots of the  statistical significances in the leptonic channels 
    discussed in Section~\ref{sec:LHCSignatures} for $30\mbox{~fb}^{-1}$ of integrated
    luminosity at the LHC.  The left-hand panel displays the
    results for the SM.  
    The right-hand panel displays the results for 
    (\(\sin\alpha=0.55\),  \(\tan\beta=3\)),  in the L2HDM.  The Standard-Model results 
    are taken from~\cite{Asai:2004ws,RichterWasNote,Belyaev:2002ua,Han:2002gp,Su:2008bj}.
    \label{fig:SigPlot}}
\end{figure}

Now that we have discussed the channels in which one might look for a 
leptonically-decaying Higgs boson at the LHC, let us investigate the 
prospects for the discovery of such a Higgs boson in the L2HDM, 
using the combined results from all channels discussed above (excepting 
the $Wh$, $Zh$ channels, which we have shown do not contribute significantly
to the discovery potential).  In particular, we focus on the region of
$\sin\alpha$ - $\tan\beta$ parameter space in which $\eta_\ell$ is large
and $\eta_q,\eta_V \sim 1$.  In this case, the cross-sections for 
processes involving a \(h\bar{\ell}\ell\) coupling are substantially 
increased, while those for processes involving \(hVV\), \(h\bar{q}q\), 
or \(hgg\) are only slightly reduced.  As before, for purposes of illustration, 
we will focus on the benchmark point ($\sin\alpha=0.55$, $\tan\beta=3$), which
exemplifies this situation nicely.  
In Fig.~\ref{fig:SigPlot}, we show the effect of the coupling-constant
modifications on the discovery potential of a light Higgs boson for 
this particular benchmark point.  In the right-hand panel, the statistical
significance associated with each of the relevant leptonic channels 
discussed in Section~\ref{sec:LHCSignatures} is displayed as a 
function of Higgs mass for our chosen
benchmark point in the L2HDM.  
The SM results for the same processes are
shown in the left-panel for comparison.  The results in each panel correspond to an integrated 
luminosity of $\mathcal{L}=30~\mbox{~fb}^{-1}$. 

It is apparent from Fig.~\ref{fig:SigPlot} that
$qq'\rightarrow qq'h(h\rightarrow\tau^+\tau^-)$ is one of the 
most promising
detection channels for the chosen benchmark point in the
L2HDM, as in the SM. 
 For this particular
choice of parameters, \(\eta_V\eta_\ell\approx 1\) and $\Gamma_{\mathit{tot}}(h)$
does not deviate drastically from $\Gamma_{\mathit{tot}}^{SM}(h)$ 
(see Fig.~\ref{fig:GamTotPlot}), and consequently the overall significance 
level in this channel is essentially unchanged from its SM value.  However,
in other regions of parameter space, drastic amplifications can occur:
for example, the choice ($\sin\alpha = 0.3$, $\tan\beta=7$) results in
a amplification of the statistical significance for the same
process by a factor of $\sim 4$.
It should also be noted that in the ($\sin\alpha = 0.55$, $\tan\beta=3$) case, 
the significance levels for both
$gg\rightarrow h\rightarrow \tau\tau$ and 
$t\bar{t}h(h\rightarrow\tau\tau)$ also exceed $5\sigma$.  The
processes in which the Higgs decays to muons are statistically less
significant, but also provide strong evidence at the $3\sigma$ 
level with $\gtrsim$ 100 ${\rm fb}^{-1}$ of integrated luminosity.  
Indeed, the evidence for such a Higgs boson would be dramatic
and unmistakable.  Furthermore, once the Higgs is observed in any of the 
muonic channels, the excellent invariant-mass resolution of the muon pairs 
can be used to determine the value of $m_h$ with a very high degree of precision.

\begin{figure}[ht!]
  \begin{center}
    \epsfxsize 2.75 truein \epsfbox{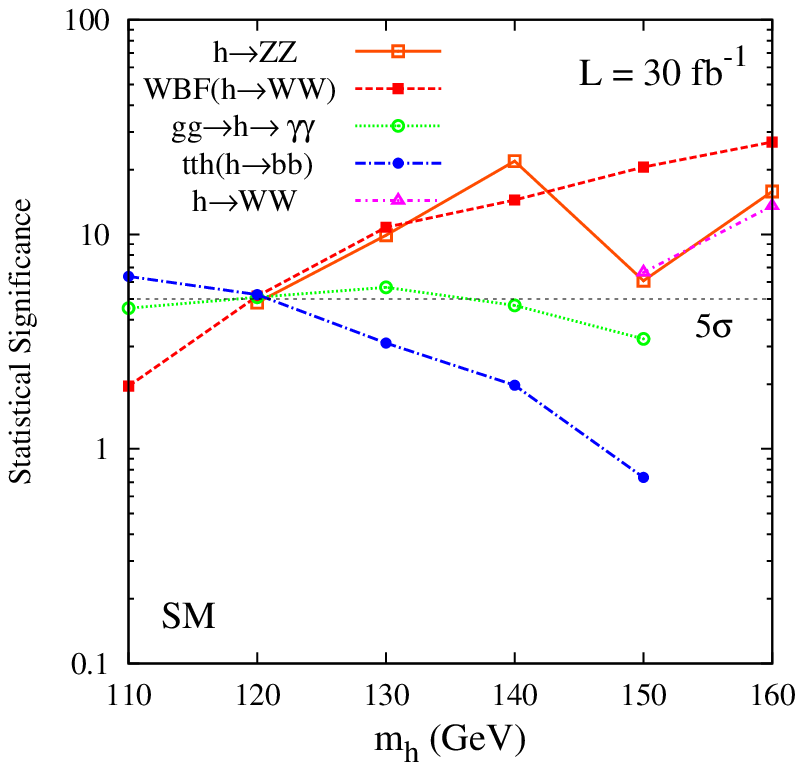}~~~~
    \epsfxsize 2.75 truein \epsfbox{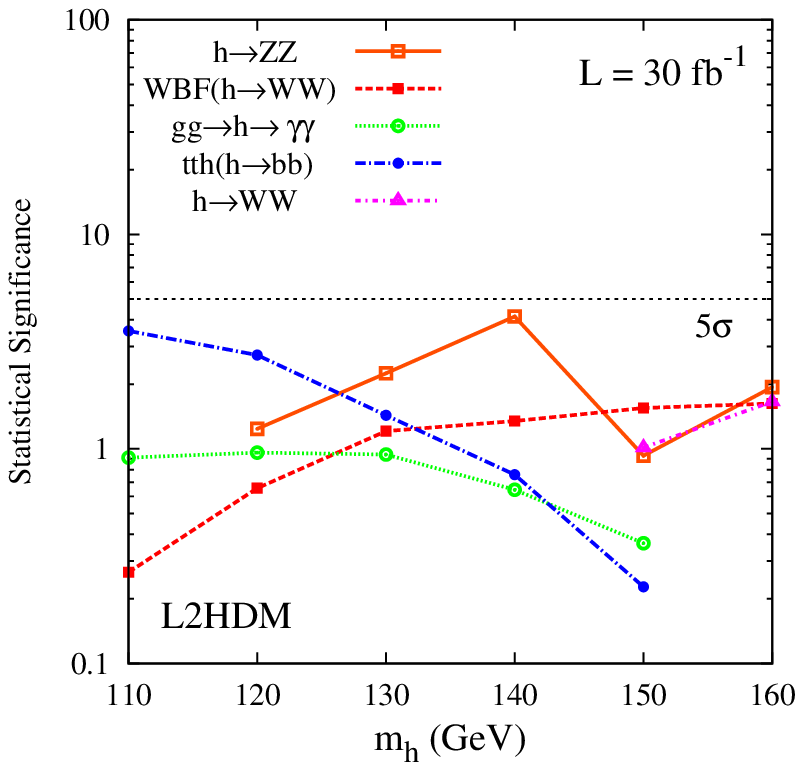}
  \end{center}
    \caption{The left-hand panel in this plot displays the statistical 
    significances in the non-leptonic channels that contribute significantly 
    to the discovery potential of a light Higgs boson in the 
    SM for $30\mbox{~fb}^{-1}$ of integrated
    luminosity at the LHC.  The right-hand panel shows 
    the corresponding significances in the L2HDM with (\(\sin\alpha=0.55\),  \(\tan\beta=3\)).
    As in Fig.~\ref{fig:SigPlot}, the Standard-Model results 
    are taken 
    from~\cite{Asai:2004ws,RichterWasNote,Belyaev:2002ua,Han:2002gp,Su:2008bj}.
    \label{fig:SigPlotOther}}
\end{figure}

\begin{figure}[ht!]
  \begin{center}
    \epsfxsize 3.5 truein \epsfbox{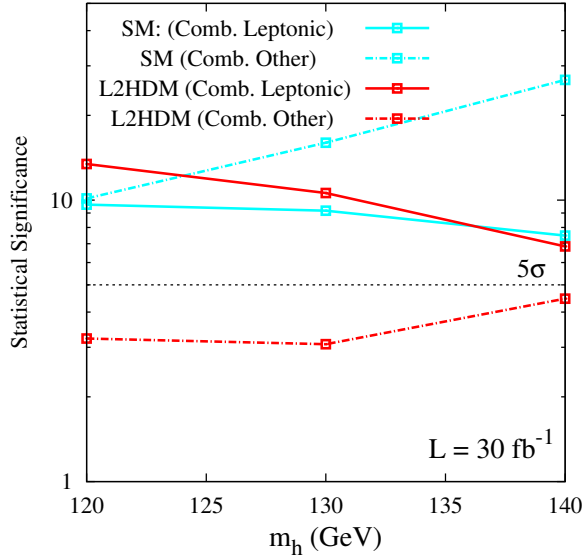}
  \end{center}
    \caption{The combined statistical significances for the leptonic 
    channels discussed in Section~\ref{sec:LHCSignatures} (solid curves), as well as
    the combined significances for all other relevant channels for light $CP$-even Higgs discovery  (dash-dotted curves)
    in the low to intermediate-mass region, both in the SM (light curves)
    and in the L2HDM (dark curves) at the benchmark point ($\sin\alpha = 0.55$, $\tan\beta=3$).  The dotted,
    horizontal line corresponds to a statistical significance at the $5\sigma$ level. 
      \label{fig:SigsComb}}
\end{figure}

While the significances in those channels which involve a leptonically-decaying Higgs 
can potentially be amplified in L2HDM, those in other channels
useful for the detection of a SM Higgs may be substantially suppressed. 
This is illustrated in Fig.~\ref{fig:SigPlotOther}, which shows the significance of
discovery in each individual channel which contributes meaningfully to the
discovery potential of a SM Higgs boson in the low to intermediate-mass region, 
both in the SM (left-hand panel)
and in the L2HDM at the benchmark point 
(\(\sin\alpha=0.55\), \(\tan\beta=3\)) (right-hand panel).  
In the latter case, there is no single, 
non-leptonic channel in which evidence for the Higgs boson can be obtained at the
$5\sigma$ level.  To further illustrate the point, 
in Fig.~\ref{fig:SigsComb}, we display the combined 
statistical significances for the leptonic channels discussed in 
Section~\ref{sec:LHCSignatures}, as well as the combined significances for all other 
relevant channels for Higgs discovery, both in the 
SM and in the L2HDM at the benchmark point 
($\sin\alpha = 0.55$, $\tan\beta=3$).   Indeed, for this particular parameter choice, all 
relevant non-leptonic channels are suppressed relative to their Standard-Model to such 
an extent that, for most of the $120\mbox{~GeV}\lesssim m_h \lesssim 140\mbox{~GeV}$
mass window displayed in the plot, their combined significance does not even provide
$3\sigma$ evidence for --- much less a $5\sigma$ discovery of --- a light Higgs boson.
On the other hand, statistical significance for leptonic Higgs decay channels are enhanced, therefore becoming the dominant discovery channels for the light $CP$-even Higgs in the L2HDM model.
This clearly illustrates the crucial role leptonic channels can play 
in the LHC phenomenology of models with extended (and particularly leptophilic) 
Higgs sectors.

We emphasize that these plots represent the results for a single benchmark point, and one
in which the $\eta$-factors are not particularly extreme.  There exist other points in
the parameter space of the model allowed by all constraints for which the deviations of
the effective couplings of $h$ to the other fields in the theory are even more severe.  As
an example, consider the case in which $\sin\alpha=0.65$ and $\tan\beta=2.2$, for which
$\eta_q=0.84$, $\eta_\ell = -1.57$, and $\eta_{W,Z}=0.30$.  For this choice of parameters,
most of the standard Higgs discovery channels --- those involving $h\rightarrow WW^\ast$ 
and $h\rightarrow ZZ^\ast$, as well as all weak-boson-fusion processes not involving
direct Higgs decays to leptons --- are strongly suppressed; furthermore, other 
contributing channels such as $gg\rightarrow h\rightarrow \gamma\gamma$ and 
$t\bar{t}h(h\rightarrow b\bar{b})$ are also moderately suppressed.  In such a case, the 
leptonic channels discussed in Section~\ref{sec:LHCSignatures} --- especially ones such as
$tth(h\rightarrow\tau\tau)$, which do not involve a direct coupling between $h$ 
and the electroweak gauge bosons --- may well constitute the only observable 
evidence of the Higgs boson, and would thus be crucial for its discovery 
at the LHC.  

\section{Conclusion\label{sec:Conclusion}}

The phenomenology of a light Higgs boson in Two-Higgs-Doublet Models can differ
drastically from that of a SM Higgs.  In this work, we have focussed on
one particularly interesting example: a leptophilic 2HDM, in which
different Higgs bosons are responsible for giving masses to the quark and lepton 
sectors.  
We have examined the effect
of such a modification on the collider phenomenology of a light Higgs boson in a
decoupling regime in which the only light scalar is a Standard-Model-like Higgs 
boson, and have
shown that a number of collider processes involving the direct decay of the Higgs 
to a pair of charged leptons can play a crucial role in its discovery.  In
particular, we have shown that there are regions of parameter space in which the 
Higgs-boson couplings to leptons can be greatly enhanced.  This can have a
potentially dramatic effect on the Higgs discovery potential, as signals involving 
direct, leptonic decays of the Higgs can be substantially amplified.  At the same time,
signals in some (or in some cases, even all) of the other conventional channels useful 
for the detection of a Standard Model Higgs boson can suffer a dramatic suppression.   
Even when coupling modifications are not severe, leptonic decay processes will also 
play an important rule in differentiating between the Higgs sector of the Standard 
Model and that of other, more complicated scenarios.

\section{Acknowledgments}

We would like to thank
Keith Dienes, Tao Han, Chung Kao and Lisa Randall 
for fruitful discussions and correspondence.
We would also like to thank the Kavli Institute of 
Theoretical Physics at Santa Barbara for its hospitality during the 
completion of this work.  This work was supported in part
by the Department of Energy under Grant~DE-FG02-04ER-41298.

Note added: After the completion of the work reported in this paper, a number of 
papers~\cite{Barger:2009me,Goh:2009wg,AokiHA} 
appeared which discuss the phenomenology of the L2HDM.  
Ref.~\cite{Barger:2009me} gives a brief presentation on the effect of 
effective-coupling modification in the decoupling regime.  Their results agree with
ours.
Refs.~\cite{Goh:2009wg} 
and~\cite{AokiHA} focussed on the situation in which $H$, $A$ and $H^\pm$ are light.

\end{document}